\documentclass[sigconf,nonacm]{acmart}

\usepackage{pvldb}
\usepackage{amsthm}
\usepackage{amsmath}
\usepackage{array}
\usepackage{algorithm}
\usepackage[noend]{algpseudocode}
\usepackage{multirow}
\usepackage{subcaption}
\usepackage{xcolor}

\setcopyright{none}
\newcommand{\system}{\textsc{Recast }}
\newcommand{\recastq}{\textsc{RecastQ }}
\newcommand{\quic}{\textsc{Quic }}

\newcommand{\Paragraph}[1]{\par\addvspace{1.5pt}\noindent\textbf{#1.}}

\newtheoremstyle{recastplain}
  {0.5\baselineskip}
  {0.5\baselineskip}
  {\itshape}
  {0pt}
  {\bfseries}
  {.}
  {0.5em}
  {}
\newtheoremstyle{recastdefinition}
  {0.5\baselineskip}
  {0.5\baselineskip}
  {\normalfont}
  {0pt}
  {\bfseries}
  {.}
  {0.5em}
  {}
\theoremstyle{recastplain}
\newtheorem{theorem}{Theorem}[section]
\newtheorem{lemma}[theorem]{Lemma}
\newtheorem{proposition}[theorem]{Proposition}

\theoremstyle{recastdefinition}
\newtheorem{definition}[theorem]{Definition}
\newtheorem{assumption}[theorem]{Assumption}

\theoremstyle{recastplain}

\makeatletter
\renewcommand{\@mkauthors@iii}{%
  \gdef\@currentauthors{}%
  \gdef\@currentaffiliation{}%
  \global\setbox\mktitle@bx=\vbox{%
    \noindent\unvbox\mktitle@bx\par\medskip
    \centering
    \begingroup
    \setlength{\tabcolsep}{3pt}
    \@affiliationfont
    \begin{tabular*}{\textwidth}{@{\extracolsep{\fill}}cccc@{}}
      {\@authorfont\href{https://orcid.org/0009-0007-1031-8708}{Zehai Yang}} &
      {\@authorfont\href{https://orcid.org/0009-0001-0333-3973}{Fengyi Tao}} &
      {\@authorfont\href{https://orcid.org/0009-0002-9276-5723}{Wenzhuo Dong}} &
      {\@authorfont\href{https://orcid.org/0009-0000-1043-6236}{Shimin Chen}\textsuperscript{*}} \\
      \href{mailto:yangzehai20z@ict.ac.cn}{yangzehai20z@ict.ac.cn} &
      \href{mailto:taofengyi20@mails.ucas.ac.cn}{taofengyi20@mails.ucas.ac.cn} &
      \href{mailto:dongwenzhuo20@mails.ucas.ac.cn}{dongwenzhuo20@mails.ucas.ac.cn} &
      \href{mailto:chensm@ict.ac.cn}{chensm@ict.ac.cn}
    \end{tabular*}\par
    \endgroup\smallskip
    \@affiliationfont
    ACS, SKLP, Institute of Computing Technology, Chinese Academy of Sciences\par
    University of Chinese Academy of Sciences, Beijing, China\par\bigskip
  }%
}
\makeatother

\begin{document}

\title{Recast: Refinement-Centered Approximate Nearest Neighbor Search Optimized for Storage}

\author{Zehai Yang}
\orcid{0009-0007-1031-8708}
\email{yangzehai20z@ict.ac.cn}
\author{Fengyi Tao}
\orcid{0009-0001-0333-3973}
\email{taofengyi20@mails.ucas.ac.cn}
\author{Wenzhuo Dong}
\orcid{0009-0002-9276-5723}
\email{dongwenzhuo20@mails.ucas.ac.cn}
\author{Shimin Chen}
\authornote{Corresponding author.}
\orcid{0009-0000-1043-6236}
\email{chensm@ict.ac.cn}
\affiliation{
  \institution{ACS, SKLP, Institute of Computing Technology, Chinese Academy of Sciences}
  \institution{University of Chinese Academy of Sciences}
  \city{Beijing}
  \country{China}
}

%\author{Zehai Yang \ \ \ \ \ Fengyi Tao \ \ \ \ \ Wenzhuo Dong \ \ \ \ \ Shimin Chen}
%\authornote{Shimin Chen is the corresponding author.}
%
%\affiliation{
%  \institution{ACS, SKLP, Institute of Computing Technology, Chinese Academy of Sciences}
%  \institution{University of Chinese Academy of Sciences}
%  \city{Beijing}
%  \country{China}
%}
%\email{\{yangzehai20z, chensm\}@ict.ac.cn}

\renewcommand{\shortauthors}{Yang et al.}

\begin{abstract}

Large-scale ANNS systems often exploit storage in addition to DRAM for
supporting vector search.  The central question is how to divide the
work between memory and storage for achieving both high query
throughput-recall and good index maintenance performance.  Existing
storage-optimized ANNS paradigms fall short of providing a complete
answer: disk-resident graph indexes have difficulty in supporting
efficient vector deletions, while posting list based systems often
incur larger amount of disk I/Os for queries.

% Large-scale ANN systems split state between memory and external storage. Disk graphs and posting indexes access external structures before all item-level decisions are complete. SIMD-accelerated quantized scanning supports an efficient IVF-style scan-and-refine pipeline when exact vectors are in memory. A natural storage extension keeps the compact scan state in memory and moves exact vectors to external storage, but this turns refinement depth into exact-vector access work. Conventional fixed-$R$ refinement cannot adapt to query-dependent demand, while uncertainty-aware selection must remain cheap enough for high-throughput scans.

In this paper, we introduce a refinement-centered paradigm for
storage-optimized ANNS systems.  We propose Recast as a concrete
design of this paradigm.  Recast stores the raw vectors on disk, and
keeps an IVF index with our proposed RecastQ quantization encoding in
memory.
For an ANNS query, Recast selects nprobe inverted lists, and performs
SIMD-based FastScan on the RecastQ codes in the lists to compute a
list of refinement candidates in memory.  Then, it performs disk I/Os
to retrieve the candidate vectors, compute exact distances, and refine
the top-$K$ result.  To reduce disk I/Os, RecastQ supports the
computation of uncertainty intervals so that the true distance falls
in the corresponding interval with high probability.  Then, we propose
a QUIC policy to select a minimal set of candidates to refine based on
the uncertainty intervals,  thereby minimizing the number of I/Os for
retrieving the candidate vectors.   
For index maintenance, both the on-disk raw vectors and the in-memory
IVF structure are amenable to insertions and deletions.

Experiments on real-world data sets show that Recast processes batched
insertions and deletions $3.10$--$37.27\times$ faster than SPFresh and
OdinANN, while its QPS at 0.99 recall changes by at most $6.9\%$ after
updating $50\%$ of the corpus. For static queries at 0.99 recall,
Recast achieves $1.56$--$2.01\times$ the QPS of the strongest baseline
with the matched memory budget on each data set, while requiring
$1.22$--$6.75\times$ less external storage than the evaluated systems.
At matched in-memory space, RecastQ also provides a better
Recall--refinement frontier than RaBitQ.

% We present Recast, a refinement-centered storage ANN index that forms an uncertainty interval for each scanned vector and selects refinement candidates against a global top-$K$ boundary. When the intervals are valid, this rule preserves the exact top-$K$ within the scanned set and is interval-only minimal. This result motivates RecastQ, a refinement-oriented encoding that combines directional PQ4 codes with compact uncertainty-related metadata; blockwise SIMD execution keeps in-memory processing efficient. Together, these techniques complete refinement candidate selection before exact-vector access, leaving external storage to supply only the selected exact vectors. This separation also enables batched list-local maintenance.

% Experiments on real-world datasets show that Recast processes phase-separated 1M-vector replacement batches 3.33$\times$ faster than SPFresh, while its fixed-configuration QPS at recall 0.99 changes by at most 4.2\% after updating 50\% of the corpus.  Recast achieves graph-competitive high-recall throughput under matched loaded-index budgets, with comparable query-ready memory but 2.9×–6.7× less external query data than the evaluated baselines.  RecastQ improves the matched-space Recall--refinement frontier over RaBitQ.

\end{abstract}


\maketitle
\begingroup
\renewcommand{\thefootnote}{}
\footnotetext{\href{https://creativecommons.org/licenses/by/4.0/}{\includegraphics[height=2.5ex]{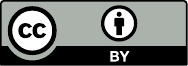}}\quad
This work is licensed under a Creative Commons Attribution 4.0 International License.}
\endgroup
\hypersetup{pdfauthor={Zehai Yang, Fengyi Tao, Wenzhuo Dong, Shimin Chen}}

\pagestyle{plain}

% Local preprint spacing; leave the class definitions and appendix untouched.
\begingroup
\makeatletter
\patchcmd{\section}{-.75\baselineskip}{-.6\baselineskip}{}{}
\patchcmd{\subsection}{-.75\baselineskip}{-.6\baselineskip}{}{}
\makeatother
\section{Introduction}
\label{sec:intro}

Approximate nearest-neighbor search (ANNS) for embedding vectors is
widely used in recommendation~\cite{col_filter_recommend}, image
retrieval~\cite{liu2007survey}, retrieval-augmented
generation~\cite{RAG}, and sparse-attention optimized KV-cache in
large language model serving~\cite{RetrievalAttention, AlayaDB}.  
For sufficiently large vector collections, it is often
uneconomical to keep all the vectors and the search structures in
DRAM.  Therefore, large-scale ANNS systems often exploit storage in
addition to DRAM for supporting vector search.  For storage-optimized
ANNS systems, the central question is how to divide the work between
memory and storage for achieving high query throughput without
sacrificing the recall and index maintenance performance.

% Approximate nearest-neighbor (ANN) search supports recommendation~\cite{col_filter_recommend}, image retrieval~\cite{liu2007survey}, retrieval-augmented generation~\cite{RAG} and selecting KV-cache entries for sparse attention~\cite{RetrievalAttention, AlayaDB}.  At hundred-million and billion-vector scale, keeping every exact vector and search structure in DRAM is often uneconomical. Storage ANN systems therefore divide state between a resident index and an external tier. The central systems question is how to divide search work between resident memory and external storage without sacrificing high-recall performance.

Existing storage-optimized ANNS systems take two influential
paradigms.  Disk-resident graph indexes, such as
DiskANN~\cite{DiskANN}, Starling~\cite{Starling}, and
OdinANN~\cite{OdinANN}, store the vectors and a proximity graph on
disk and exploit memory for keeping part of the search structures and
caching recent entries to accelerate graph traversal.  Posting list
based systems, such as SPANN~\cite{SPANN} and SPFresh~\cite{SPFresh},
store each vector into (possibly multiple) posting lists on disk and
use an in-memory index to select and retrieve postings for subsequent
distance evaluation.  While both paradigms have shown good storage
ANNS performance, neither provides a complete answer to the above
question: graph-based solutions have difficulty in supporting
efficient vector deletions, and posting list based systems often incur
larger amount of disk I/Os for queries.  

% Existing systems occupy two influential design points. Disk-resident graph indexes read graph records whose contents guide later traversal.  Representative systems include DiskANN~\cite{DiskANN}, Starling~\cite{Starling}, and OdinANN~\cite{OdinANN}. Partition systems such as SPANN~\cite{SPANN} and SPFresh~\cite{SPFresh} select partitions in memory and retrieve posting-list vectors for subsequent distance evaluation. Both families have demonstrated strong storage ANN performance. They also show that storage ANN systems can assign fundamentally different roles to the external tier.

% These alternatives motivate a broader question: can a storage ANN system complete refinement candidate selection from compact resident information before exact-vector access?

In this paper, we would like to rethink the design of
storage-optimized ANNS systems.  One natural way is to revisit a
widely used main-memory ANNS paradigm: IVF PQ with FastScan and
Refinement.
IVF narrows the search to a subset of inverted lists, vector
quantization keeps per-vector state compact, and PQ FastScan
accelerates distance computation on compressed codes with SIMD
operations~\cite{IVF,PQ,PQfs}.
Given the short candidate list computed by IVF PQ with FastScan,
Refinement then accesses a small number of candidate vectors and
computes exact distances to obtain highly accurate top-$K$ results.
Together, these features form an effective scan-and-refine paradigm
that achieves high query throughput-recall performance.  At the same
time, the IVF structure is known to be amenable to index maintenance.

%One natural way to pursue this question is to revisit a high-throughput resident-memory ANN pattern. IVF narrows the search to a subset of lists, vector quantization keeps per-vector state compact, and SIMD-oriented techniques such as PQ FastScan evaluate compressed codes efficiently~\cite{IVF,PQ,PQfs}. The pipeline then accesses exact vectors for a shortlist, computes exact distances, and refines the final result. Together, these stages form an efficient scan-and-refine organization when compact codes and exact vectors are resident.

Inspired by the main-memory IVF PQ with FastScan and Refinement, we
propose \system (Refinement-Centered Approximate nearest neighbor
search optimized for STorage).
\system stores the exact vectors on disk, and keeps the IVF structure
and quantization codes in memory.  Note that quantization can
effectively reduce the memory footprint to a small fraction (e.g.,
1/4--1/10) of the original data set size.  For an ANNS query, \system
selects \texttt{nprobe} inverted lists whose centroids are closest to
the query vector, and scans the quantization codes in the
\texttt{nprobe} lists to compute the list of refinement candidates in
memory.  Then, \system performs disk I/Os to retrieve the candidate
vectors, compute exact distances, and refine the top-$K$ result.
\system brings the following benefits:

\begin{list}{\labelitemi}{\setlength{\leftmargin}{3.5mm}\setlength{\itemindent}{0mm}\setlength{\topsep}{0.5mm}\setlength{\itemsep}{0mm}\setlength{\parsep}{0.5mm}}

\item {\bf Benefit 1: Efficient index maintenance}.  Both the on-disk
and in-memory structures of \system can efficiently support index
maintenance.  
For on-disk structures, \system only performs vector-record and 
metadata updates for vector insertions and deletions, without 
maintaining graph connectivity or replicated postings. In comparison, the existing
paradigms have to either modify the graph structure or update
potentially multiple postings for each updated vector.
For in-memory structures, insertions and deletions can also be easily
handled in the IVF structure of \system.

\item {\bf Benefit 2: Independent I/O operations}.  Unlike the
dependent I/Os due to graph traversals in graph-based solutions,
\system issues independent disk I/Os for the candidate vectors.
Therefore, it is straight-forward to optimize disk accesses with
asynchronous I/Os, and deduplicate and merge the I/O requests for
vectors on the same disk page.

\end{list}
However, \system faces and addresses two important challenges:
\begin{list}{\labelitemi}{\setlength{\leftmargin}{3.5mm}\setlength{\itemindent}{0mm}\setlength{\topsep}{0.5mm}\setlength{\itemsep}{0mm}\setlength{\parsep}{0.5mm}}

\item {\bf Challenge 1: Minimizing the number of disk I/Os}.  
Compared to the refinement in the main-memory IVF index, the
refinement in \system performs disk I/Os, incurring drastically higher
cost.  Consequently, it is important to minimize the number of
candidates to refine.

\hspace{2mm} 
The conventional fixed-$R$ policy selects the top-$R$ candidates which
are closest to the query vector.  However, a fixed $R$ can hardly
capture the difference across queries: a large $R$ incurs wasteful
disk I/Os on easy queries, whereas a small $R$ downgrades the recall
on difficult queries.
Figure~\ref{fig:oracle-depth} shows that the per-query optimal $R_q^*$
for obtaining all top-10 results can be orders of magnitude different
across queries on SIFT100M and DEEP100M data sets with PQ and OPQ
encodings.  

% Figure~\ref{fig:fixed-r-breakdown} shows that for achieving over 99\% recalls, vector I/Os account for 94.1--96.4\% of the measured fixed-$R$ latency.

% \begin{figure}[t]
%  \centering
%  \begin{subfigure}[t]{0.49\linewidth}
%    \centering
%    \includegraphics[width=\linewidth]{figures/oracle_refinement_depth_cdf.pdf}
%    \caption{Oracle refinement-depth CDF.}
%    \label{fig:oracle-depth}
%  \end{subfigure}
%  \hfill
%  \begin{subfigure}[t]{0.49\linewidth}
%    \centering
%    \includegraphics[width=\linewidth]{figures/fixed_refinement_latency_breakdown.pdf}
%    \caption{Fixed-$R$ latency breakdown.}
%    \label{fig:fixed-r-breakdown}
%  \end{subfigure}
%  \caption{Fixed-$R$ refinement mismatches query demand under external storage.}
%  \label{fig:motivation-refinement}
%\end{figure}

\hspace{2mm} 
To address this issue, we propose \recastq encoding and \quic
(Query-aware Uncertainty-guIded Candidate selection).  \recastq
enables \system to associate the estimated distance $d_i$ for every
vector $i$ with an uncertainty interval $[L_i, U_i]$ so that $d_i \in
[L_i, U_i]$ with high probability.  In this way, \quic can select a
minimal set of candidates that contain the top-$K$ neighbors for a
query with high confidence.  Moreover, \recastq performs
bit-gain-aware grouping, which assigns higher encoding bit budget to
vector dimensions with higher variance for improving the encoding
quality and reducing the uncertainty interval size.

\item {\bf Challenge 2: Reducing in-memory computation cost}.  After
optimizing the disk I/O accesses, we find that the cost of in-memory
IVF computation sometimes surpasses the I/O cost.  The in-memory
search consists of (1) selecting \texttt{nprobe} IVF lists, which
computes the distances of all list centroids, and (2) scanning the
\recastq codes in the chosen lists.  For (1), we encode the list
centroids in PQ and employ FastScan to accelerate the distance
computation for centroids.  For (2), we implement FastScan on the
\recastq codes.  We design the layout of \recastq to be compatible
with that of PQ4 FastScan, which stores codes in 32-vector packed
blocks.  Unlike the int8 operations in PQ FastScan, we perform 32-bit
floating point SIMD operations to improve the accuracy of the distance
computation.

\end{list}

Experiments on real-world data sets show that \system processes
batched insertions and deletions $3.10$--$37.27\times$ faster than
SPFresh and OdinANN, while its QPS at 0.99 recall changes by at most
$6.9\%$ after updating $50\%$ of the corpus. For static queries at
0.99 recall, \system achieves $1.56$--$2.01\times$ the QPS of the
strongest baseline with the matched memory budget on each data set,
while requiring $1.22$--$6.75\times$ less external storage than the
evaluated systems.  At matched in-memory space, \recastq also provides
a better Recall--refinement frontier than RaBitQ.

\begin{figure}[t]
    \centering
    \includegraphics[width=0.75\linewidth]{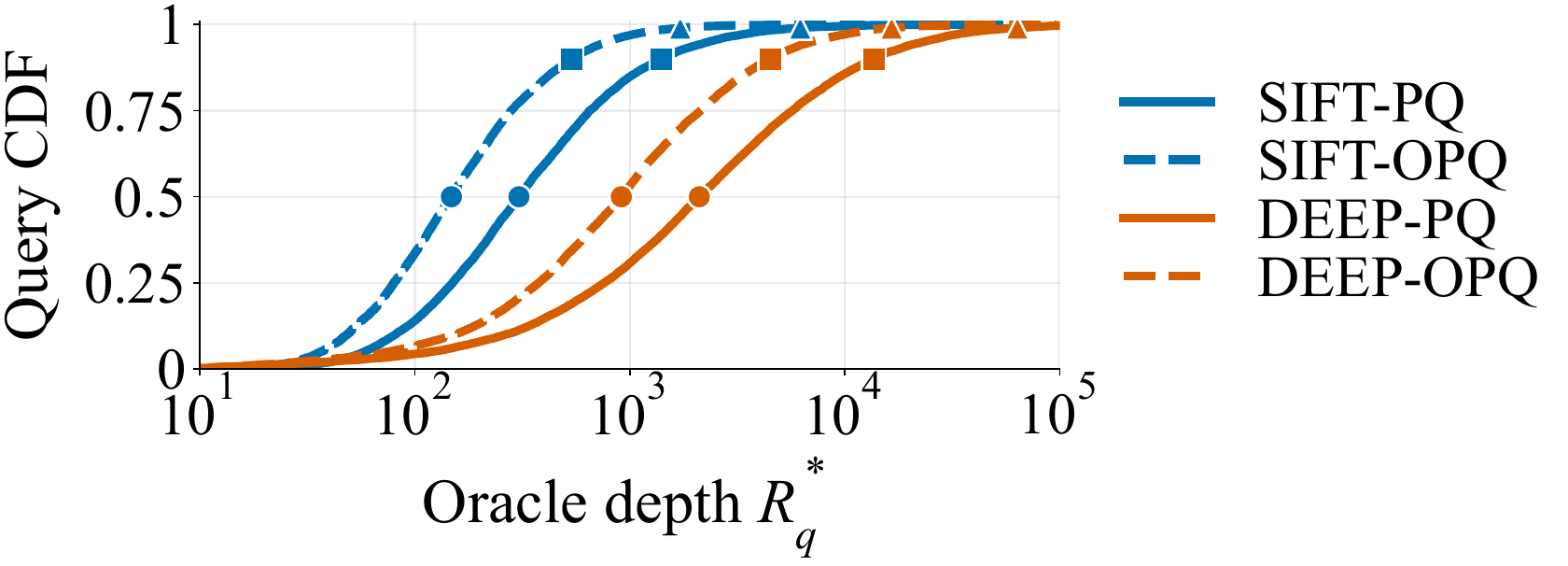}
    \vspace{-0.15in}
    \caption{Distribution of per-query optimal $R_q^*$ for obtaining
all top-10 results under the fixed-$R$ refinement policy.}
    \label{fig:oracle-depth}
    \vspace{-0.2in}
\end{figure}

The paper makes the following contributions:

\begin{list}{\labelitemi}{\setlength{\leftmargin}{3.5mm}\setlength{\itemindent}{0mm}\setlength{\topsep}{0.5mm}\setlength{\itemsep}{0mm}\setlength{\parsep}{0.5mm}}

  \item To our knowledge, we are the first to introduce the
refinement-centered paradigm for storage-optimized ANNS systems, which
provides a promising answer to the central question for achieving good
query throughput, recall, and index maintenance performance.  We
propose \system as a concrete design of this paradigm.

  \item We propose \recastq encoding, which supports uncertainty
interval computation and fast distance estimation, and \quic policy,
which for each query minimizes the set of refinement candidates to
reduce the I/O overhead.

  \item We build an end-to-end \system system.  Our experiments
comparing \system with state-of-the-art storage-optimized ANNS systems
show that \system achieves 1.56-2.01$\times$ the query throughput of
the strongest baseline at 0.99 recall, and shows strong performance
stability under dynamic workloads.

\end{list}

% We introduce a refinement-centered storage architecture and an uncertainty-guided selector that complete refinement candidate selection before exact-vector access. Under valid intervals, the selector conditionally preserves scannedset top-k and is interval-only minimal.
% 
% We design RecastQ, a refinement-oriented encoding combining directional PQ4 codes, compact uncertainty metadata, and variance-aware uneven grouping to reduce refinement demand under practical code budgets.
% 
% We build an end-to-end system with FastScan-compatible SIMD block scanning, page-aware access planning, asyn- chronous exact-vector access, and parallel batched list-local maintenance. Under the evaluated protocol, it attains competitive high-recall throughput with comparable query-ready memory, substantially less external query data, and strong stability under updates.

\begin{table*}[!t]

\caption{Comparing storage-optimized ANNS paradigms.}
\label{tab:storage-ann-design-space}
\vspace{-0.1in}

\centering
\renewcommand{\arraystretch}{1.12}
\setlength{\tabcolsep}{1.5pt}
\renewcommand{\aboverulesep}{0ex} % 0.4ex by default
\renewcommand{\belowrulesep}{0ex} % 0.65ex by default
\begin{tabular}{>{\raggedright\arraybackslash}m{3.1cm}|
                >{\centering\arraybackslash}m{3.6cm}|
                >{\centering\arraybackslash}m{3.0cm} 
                >{\centering\arraybackslash}m{2.3cm} 
                >{\centering\arraybackslash}m{1.6cm}|
                >{\centering\arraybackslash}m{1.8cm}|
                >{\centering\arraybackslash}m{1.4cm}}
\toprule
\shortstack{\textbf{Paradigm}} &
\shortstack{\textbf{Representative}\\\textbf{systems}} &
\shortstack{\textbf{Query performance}\\\textbf{(w/ same mem size)}} &
\shortstack{\textbf{Amount of disk}\\\textbf{data accessed}} &
\shortstack{\textbf{Dependent}\\\textbf{I/Os}} &
\shortstack{\textbf{Maintenance}\\\textbf{support}} &
\shortstack{\textbf{Storage}\\\textbf{footprint}} \\
\midrule
Static disk graph & DiskANN~\cite{DiskANN}, Starling~\cite{Starling} & 
High          &Low          & Yes         & -               & Moderate \\
Dynamic disk graph & \mbox{OdinANN}~\cite{OdinANN} &
Moderate      &High         & Yes         & Moderate        & Moderate \\
Static posting list & SPANN~\cite{SPANN} & 
Moderate      &High         & No          & -               & High \\
Dynamic posting list & SPFresh~\cite{SPFresh} & 
Moderate      &High         & No          & Moderate        & High \\
\midrule
\textbf{Refinement-centered} & \textbf{Recast} & 
\textbf{High} &\textbf{Low} & \textbf{No} & \textbf{Strong} & \textbf{Low} \\
\bottomrule
\end{tabular}
\vspace{-0.1in}
\end{table*}

\section{Storage-Optimized ANNS Paradigms}
\label{sec:background}

We compare storage-optimized ANNS paradigms in this section.
Table~\ref{tab:storage-ann-design-space} summarizes the comparison.
For fair comparison, we assume the same allocated memory budget across
different systems, which is used by in-memory structures for both
query processing and update support.  

We consider five aspects of the design in the table, including three
aspects on ANNS query processing, one aspect on index maintenance, and
one aspect on the space cost.
(1) \emph{Query performance} summarizes the ANNS throughput-recall
performance.  
(2) \emph{Amount of disk data accessed} accounts for not only raw
vectors but also auxiliary structures retrieved from disk during query
processing.  
(3) \emph{Dependent I/Os} indicates whether an I/O read decision can
depend on information returned by previous I/Os in the same query,
which may have adverse impact on I/O efficiency.
(4) \emph{Maintenance support} qualitatively describes whether and how
efficient the systems support vector insertions and deletions.
(5) \emph{Storage footprint} compares the total amount of storage
space consumed by vector data and auxiliary structures required by
queries and updates.

% rather than implying identical update semantics across systems.

In the following, Section~\ref{subsec:graph} and~\ref{subsec:postlist}
review the existing paradigms, then
Section~\ref{subsec:refinement-centered} introduce the
refinement-centered paradigm.

\subsection{Disk-Resident Graph Indexes}
\label{subsec:graph}

\Paragraph{Static disk-resident graphs}
DiskANN~\cite{DiskANN} is the first static disk-resident solution.  It
stores vertices (i.e., exact vectors along with graph adjacency lists)
on disk, while keeping compressed PQ codes and preserving a vertex
cache in memory.  
Starling~\cite{Starling} constructs a sampling-based in-memory
navigation graph to get better entry points for the disk graph
search, and improves I/O locality by placing vertices and
out-neighbors in the same disk blocks as much as possible.
MARGO~\cite{MARGO} further improves the disk data layout by
prioritizing out-edges based on their importance in the graph search.
GPS~\cite{GPS} appends out-neighbors' adjacency lists to each vertex
entry on disk so that an I/O obtains more graph structure information.
PipeANN~\cite{PipeANN} gives up the strict compute-I/O order across
search steps and issues asynchronous I/Os based on the current
candidate pool, regardless of uncompleted I/Os or computation.

\Paragraph{Dynamic disk-resident graphs}
For updates, FreshDiskANN~\cite{FreshDiskANN} follows a main+delta
design.  It constructs one or more TempIndexes for insertions and a
DeleteList in memory, in addition to the read-only disk-resident
graph.  When the memory is full, FreshDiskANN merges the three
structures to compute a new disk-resident graph.
IP-DiskANN~\cite{IPDiskANN} proposes an in-place deletion algorithm,
which approximates the in-neighbors of a deleted vertex by vertices
visited during the graph search, but has to periodically scan the full
index to remove dangling edges.
OdinANN~\cite{OdinANN} performs direct vector insertions to reduce
foreground performance variation during updates.  For deletions, it
buffers the deleted vector IDs in memory and merges the deletes to the
disk graph periodically.

\Paragraph{Representative systems and performance aspects}
In this paper, we consider DiskANN~\cite{DiskANN} and
Starling~\cite{Starling} as two classical solutions, and
OdinANN~\cite{OdinANN} as the state-of-the-art solution in the
paradigm of disk-resident graphs.
As shown in Table~\ref{tab:storage-ann-design-space}, the graph-based
solutions store adjacency lists with vectors on disk, leading to a
moderate storage footprint.  Moreover, there are dependency among I/O
operations during graph search.  Furthermore, OdinANN achieves
moderate index update performance.  However, compared to the static
solutions, OdinANN allocates large per-thread I/O buffers for
improving I/O efficiency, and keeps significant metadata, such as
vector mapping and allocation structures, in memory.  Given the same
memory budget, OdinANN has to reduce the PQ code length, which leads
to less accurate distance estimation, incurring higher amount of disk
I/Os and lower query performance.

\subsection{Posting List Based Systems}
\label{subsec:postlist}

\Paragraph{Static posting list based systems}
SPANN~\cite{SPANN} performs hierarchical clustering to generate a
large number of posting lists so that each posting fits into a small
number of disk blocks (e.g., 10--100KB).  Then, it assigns (boundary)
vectors to multiple postings for better recall, and constructs an
in-memory index on the medoids (i.e., vectors closest to the centroids
of postings).  A query first searches the in-memory index to obtain
candidate postings, then reads the candidate postings to compute the
top-$K$ result.

% SPANN keeps a head index in memory and stores replicated posting lists on external storage. A query selects partitions through the head index, reads their postings, evaluates the returned vectors, and reranks the best results. This organization regularizes storage access but reads posting-level data before item-level distance evaluation is complete. The in-memory head index determines the posting set before these reads, so the contents of one posting do not determine the next in-memory search step or the address of another posting.

\Paragraph{Dynamic posting list based systems}
SPFresh~\cite{SPFresh} adds incremental in-place update support to
SPANN.  Vector updates are handled locally in posting lists.  If a
posting exceeds an upper threshold, SPFresh splits it into two smaller
ones.  If the size of a posting is smaller than a lower threshold,
SPFresh deletes it and reassigns its vectors to nearby postings.

% It applies local vector reassignment and posting split/merge operations to limit index-quality degradation as the corpus changes. Updates modify both list membership and the external posting layout.

\Paragraph{Representative systems and performance aspects}
In this paper, we evaluate SPANN~\cite{SPANN} and
SPFresh~\cite{SPFresh} as two representative solutions in the paradigm
of posting list based systems.
As shown in Table~\ref{tab:storage-ann-design-space}, they incur high
storage cost because vectors can be stored redundantly in multiple
postings.  Moreover, the I/Os for retrieving postings are independent
of each other in a query.  Furthermore, since full postings of vectors
are retrieved, SPANN and SPFresh read higher amount of disk data than
graph based solutions, which can adversely impact the query
performance.

\subsection{This Paper: Refinement-Centered Paradigm}
\label{subsec:refinement-centered}

We introduce a refinement-centered paradigm for storage-optimized ANNS
systems.  As shown in Table~\ref{tab:storage-ann-design-space}, it
achieves a low storage footprint for storing vectors without
per-vector adjacency lists or redundant assignment.  Moreover, it
keeps an IVF structure on compressed vector encodings in memory.  Both
the in-memory structure and the on-disk vectors are update friendly.
Furthermore, a query searches the in-memory IVF index to obtain a list
of refinement candidates, then issues selective, independent I/Os for
retrieving the candidate vectors for refinement.  
Finally, as will be described in the \system design, the \recastq encoding and the \quic
policy minimize the number of candidate vectors and the amount of disk
I/Os, thereby improving the query performance.

\begin{figure}
  \centering
  \includegraphics[width=\linewidth]{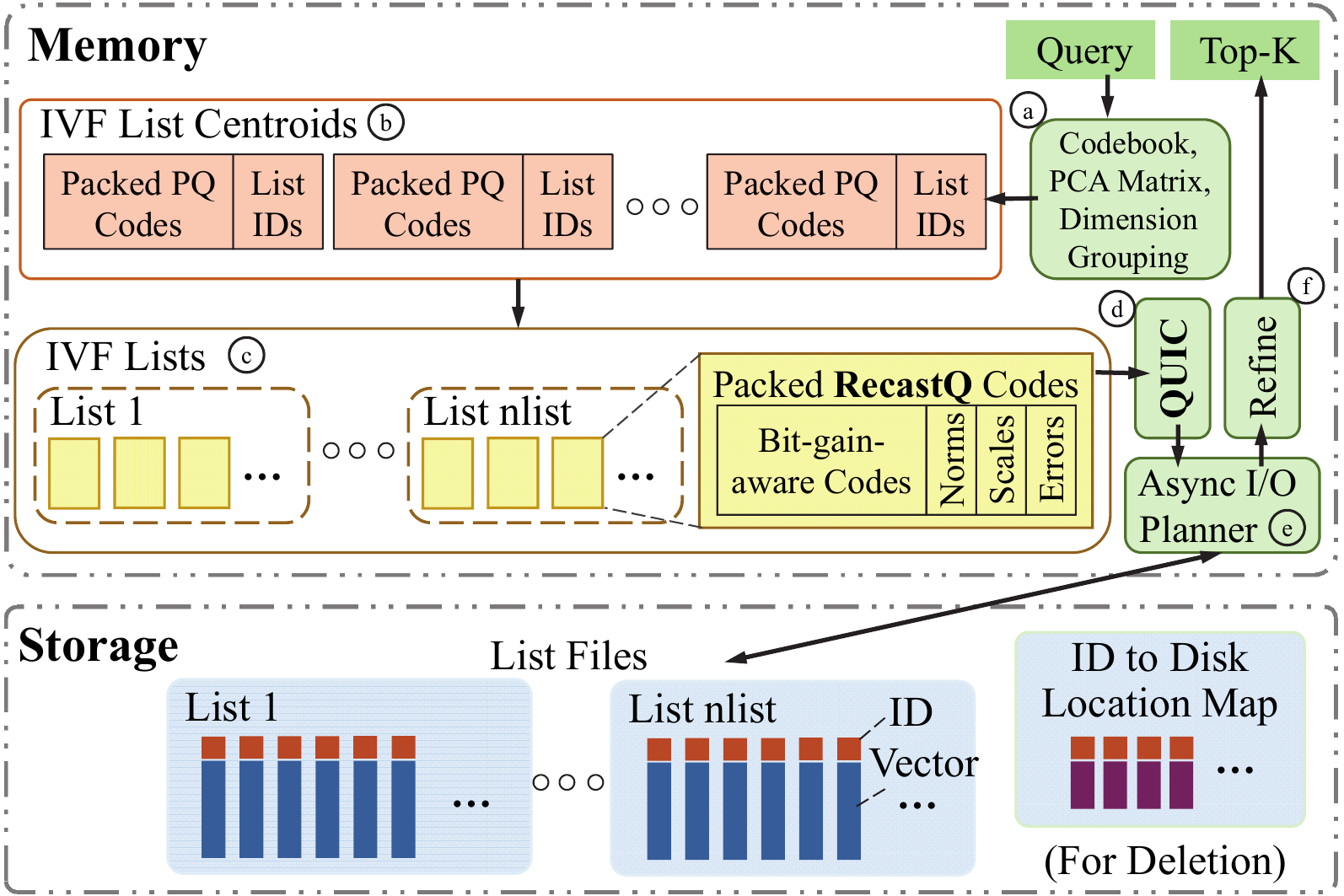}
  \vspace{-0.30in}
  \caption{\system architecture and query data flow.}
  \label{fig:architecture}
  \vspace{-0.15in}
\end{figure}

\section{Recast Overview}
\label{sec:architecture}

We propose \system (Refinement-Centered Approximate nearest neighbor
search optimized for STorage) as a concrete design of the
refinement-centered paradigm.  In the following,
Section~\ref{subsec:arch} overviews the \system architecture.  Then,
Section~\ref{subsec:anns} and~\ref{subsec:maintenance} describe its
ANNS query and update support, respectively.

% This section describes the index and query pipeline that enforce the constraint; Sections~\ref{sec:refinement} and~\ref{sec:encoding} develop its selector and encoding, while Section~\ref{sec:memory-optimizations} describes the in-memory execution path.

\subsection{Recast Architecture}
\label{subsec:arch}

Figure~\ref{fig:architecture} depicts the \system architecture.  We
describe the in-memory and on-disk structures, then analyze the space
cost.

\Paragraph{In-memory structures}
\system constructs an IVF index in memory.  It is composed of a list
of centroids and the IVF lists.

First, list centroids are encoded in PQ codes for saving space and
reducing computation cost.  The same PQ quantizer is used for
build-time list assignment, post-build insertion, and query-time IVF
list selection.  Let $D$ be the number of vector dimensions.  We set
the codes to use $D$ PQ groups, each taking 4 bits.  Suppose each
dimension is a 32-bit floating point number.  Then, this reduces the
centroid size from $4D$ to $0.5D$ bytes, an $8\times$ space saving.
Moreover, PQ FastScan delivers $4.5\times$ speedup for the IVF list
selection.

Second, each IVF list stores \recastq codes packed in Faiss FastScan
blocks.  A \recastq code consists of $M$ code groups, each taking 4
bits, and three bytes, which encode the \emph{norm}, the \emph{scale},
and the \emph{error} (cf.  Section~\ref{subsec:metadata}).
Hence, each vector's \recastq code takes $0.5M+3$ bytes.  Note that we
do not store any vector IDs or disk block addresses in memory because
a vector's disk location can be computed from its IVF list ID and
in-list offset.  For each vector, this saves at least 4B of memory
space.

% Padding lanes have zero codes and sentinel metadata and never become results. 

\Paragraph{On-disk structures}
First, \system creates a list file for each in-memory IVF list.  A
list file stores the vector records in the same order as the
associated IVF list.  A vector record consists of a 32-bit vector ID
and the exact vector.  Suppose the disk page size is $B$ and the
vector record size is $S$.  Each page contains $n_B = \lfloor
\tfrac{B}{S} \rfloor$ records.  Hence, for the $j$-th \recastq code in
IVF list $i$, the associated vector record is the ($j$ mod $n_B$)
records in page $\lfloor\tfrac{j}{n_B}\rfloor$ in list file $i$.

% SIFT uses 132-byte records, 31 per 4-KiB page.

Second, \system also maintains an ID-to-disk-location map.  It is not
accessed in ANNS queries, but used only for deletions.

Finally, to support quick system restart, it is optional to store a
backup copy of the in-memory structures on disk.  (This is not
depicted in Figure~\ref{fig:architecture}.)  Note that if the backup
copy does not exist, one can always rebuild the in-memory IVF index
from the list files.

\Paragraph{Space cost analysis}
For the memory space, suppose there are $N$ vectors.  The IVF lists
take $N (0.5M+3)$ bytes.  Since a list ID takes 4B, the IVF list
centroids take $nlist (0.5D+ 4)$ bytes.  In total, the in-memory IVF
index takes $nlist (0.5D + 4) + N (0.5M+3)$ bytes.  For a given memory
budget, $M$ can be easily derived.

For the disk footprint, the list files take about $\tfrac{N}{n_B}B$
bytes, and the ID-to-disk-location map takes $8N$ bytes.  In total,
the on-disk structures take $\tfrac{N}{\lfloor B/S \rfloor}B + 8N$
bytes.  Since the dimensions of the vector data sets are often on the
order of 100--1000, $S \gg 8$.  Therefore, the disk footprint of
\system is often quite close to $NS$, the space required for storing
the raw vectors.

\begin{algorithm}[t]
\caption{\system ANNS algorithm.}
\label{alg:recast-search}
\begin{algorithmic}[1]
\Require query $q$, output size $K$, number of lists to probe
$nprobe$, uncertainty interval scale $z$, I/O queue depth $Q$
\State $\mathcal{L}\gets\textsc{SelectIVFLists}(q,nprobe)$
\State $\mathcal{C},\theta_K\gets\textsc{ScanWithQUIC}(q,\mathcal{L},K,z)$
%\State $\mathcal{P}\gets\textsc{GetListFilePageIDs}(\mathcal{C})$
\State $Result \gets\textsc{AIOReadAndRefine}(q,\mathcal{C},K,Q)$
\State \Return $Result$
\end{algorithmic}
\end{algorithm}

\subsection{ANNS Algorithm}
\label{subsec:anns}

Algorithm~\ref{alg:recast-search} lists the \system ANNS algorithm.
The query data flow is illustrated in Figure~\ref{fig:architecture}.
First, given a query, \system selects $nprobe$ IVF lists by performing
PQ FastScan on the packed centroid codes (Line 1, Step (a) and (b)).
Second, it scans the packed \recastq codes in the selected IVF lists
to compute the estimated distance and the uncertainty interval for
each item in the lists, while the refinement selector applies the
\quic policy (Line 2, Step (c) and (d)).  Specifically, the selector
computes a top-$K$ upper bound $\theta_K$, and obtains a minimal set
of refinement candidates, whose lower bounds of the uncertainty
intervals $\leq \theta_K$ (cf. Sections~\ref{sec:refinement}).
Finally, \system derives the disk locations of the refinement
candidates, plans and issues asynchronous I/Os, and computes accurate
vector distances for refinement (Line 3, Step(e) and (f)).  The vector
IDs in the returned vector records are used to produce the top-$K$
result.

% The query proceeds through coarse list selection, compressed-code scanning and uncertainty estimation, refinement candidate selection, asynchronous I/O planning, exact-vector access, exact-distance computation, and final refinement, in that order.

\Paragraph{Optimizing I/O Operations for Retrieving Vectors}
Given the set $\mathcal{C}$ of refinement candidates, \system computes
the disk locations of the $|\mathcal{C}|$ vectors (cf.
Section~\ref{subsec:arch}).  Then, it sorts (list file ID, disk page
ID)s and merges items that are mapped to the same disk page or
adjacent disk pages into a single I/O request.  Given the queue depth
$Q$, a query worker submits $Q$ I/O requests initially.  Whenever
submitted I/O requests complete, it refills the released slots with
remaining requests.  In this way, \system exploits the independent
nature of the I/O requests and the parallelism of the underlying
storage devices (e.g., SSDs) for improving I/O efficiency.

% Multiple refinement candidates on one page therefore trigger one page read, while exact distance is still evaluated once per candidate.

% The involved list files are opened once per query and share the query worker's asynchronous I/O context.

\Paragraph{Cost Analysis}
First, \textsc{SelectIVFLists} takes $O(nlist \cdot D)$ work.
Second, \textsc{ScanWithQUIC} scans an average
$\tfrac{nprobe}{nlist}N$ \recastq codes, computes the top-$K$ upper
bound $\theta_K$ using an $O(K)$-sized heap, and checks the
uncertainty lower bound of each item against $\theta_K$.  
Third, \textsc{AIOReadAndRefine} takes $O(|C|\log |C|)$ work for the
sorting and planning, issues up to $|C|$ asynchronous I/O requests,
computes $|C|$ full distances, and uses a heap of size $O(K)$ to
obtain the top-$K$ result.  
In summary, for an ANNS query, \system takes $O(nlist \cdot D+
\tfrac{nprobe}{nlist}N (M + \log K) + |C|(\log |C|+ \log K + D))$ computation
work, and performs $O(|C|)$ I/O accesses.

% For $S_{\mathrm{pad}}$ padded scanned rows, $M$ PQ4 groups, and $C$ refinement candidates, packed scanning and uncertainty arithmetic perform $O(S_{\mathrm{pad}}M)$ scalar-equivalent work in SIMD blocks.  The global boundary uses $O(K)$ state and at most $O(S\log K)$ heap work, while page planning costs $O(C\log C)$ time and $O(C+E)$ temporary state for $E$ merged extents.

\begin{algorithm}[t]
\caption{Insertion and deletion algorithms.}
\label{alg:maintenance}
\begin{algorithmic}[1]
\Procedure{InsertBatch}{$X$}
    \State $ID_X \gets \Call{ReserveVectorIDs}{|X|}$
    \State $L_X \gets \Call{AssignLists}{X}$
    \State $(Code_X,U_X) \gets \Call{EncodeBatch}{X}$
    \State \parbox[t]{\dimexpr\linewidth-\algorithmicindent}{
       \strut\hangindent=3em \hangafter=1 
       $(G,ID_G,X_G,Code_G,U_G) \gets$ \\
       $\Call{GroupInsertsByList}{L_X,ID_X,X,Code_X,U_X}$
    }
    \For{\textbf{each} touched list $\ell \in G$ \textbf{in parallel}}
        \State \Call{AppendPackedRecordsInMem}{$\ell,Code_G[\ell],U_G[\ell]$}
        \State $A_G[\ell] \gets$ \Call{AppendRecordsOnDisk}{$\ell,ID_G[\ell],X_G[\ell]$}
    \EndFor
    \State \Call{AddInIDtoDiskLocMap}{$ID_G,A_G$}
    \State \Return $ID_X$
\EndProcedure
\vspace{0.05in}
\Procedure{DeleteBatch}{$ID$}
    \State $A \gets \Call{LookupLocations}{ID}$
    \State \Call{ValidateDeletes}{$A$}
    \State $(G, A_G) \gets \Call{GroupDeletesByList}{ID,A}$
    \For{\textbf{each} touched list $\ell \in G$  \textbf{in parallel}}
        \State $P[\ell] \gets \Call{PlanSwapWithLast}{\ell,G[\ell]}$
        \State \Call{MovePackedRecordsInMem}{$\ell,P$}
        \State \Call{MoveRecordsOnDisk}{$\ell,P$}
    \EndFor
    \State \Call{UpdateMovedLocAndMarkTombstones}{$P$}
\EndProcedure
\end{algorithmic}
\end{algorithm}

\subsection{Index Maintenance}
\label{subsec:maintenance}

The index insertion and deletion algorithms are shown in
Algorithm~\ref{alg:maintenance}.  Note that the index structures are
updated in place, avoiding the cost of background maintenance tasks
used in previous solutions~\cite{FreshDiskANN,OdinANN}.  Moreover, for
a batch of requests, both insertion and deletion procedures divide the
requests based on the relevant lists, and perform the update
operations of each list in parallel.

For insertions, \system reserves the vector IDs for the input vectors
(Line 2), assigns the vectors to the IVF lists with the closest
centroids (Line 3), and encodes the vectors into RecastQ codes (Line
4).  Then, it groups the insert requests by the assigned lists (Line
5), and enters a parallel for loop (Line 6--8).  Each parallel
iteration processes a touched list, appending the inserted RecastQ
codes to the in-memory IVF list, and appending the inserted vector
records to the list file on disk.  Finally, the IDs and the disk
locations of the inserted vectors are added into the
ID-to-disk-location map.

% Touched lists are processed in parallel; each in-memory list receives one grouped append that fills unused lanes in its tail block and packs only appended lanes and new blocks, avoiding a per-vector unpack/repack cycle.

For deletions, given a set of vector IDs to delete, \system retrieves
their disk locations from the ID-to-disk-location map on disk (Line
12), and ensures that the disk locations are valid (i.e., not
tombstones) (Line 13).  Then, it groups the resulting locations by the
relevant lists (Line 14), and enters a parallel for loop (Line
15--18).  Each parallel iteration processes a touched list. 
For a list containing $N_l$ vectors, if there are $d_l$ deletions, the
list size after deletion is $N_l'=N_l-d_l$.  \textsc{PlanSwapWithLast}
considers the deleted slots in slot $[0, N_l')$ as target holes, and
regards the entries in $[N_l',N_l)$ that are not deleted as source
entries. It pairs the source entries with the target holes.
Subsequently, \textsc{MovePackedRecordsInMem} and
\textsc{MoveRecordsOnDisk} overwrite the target holes with source
entries, and truncate the list.  Note that for the in-memory IVF list,
overwriting an entry involves unpacking the packed code block,
updating the entry, and re-packing the code block.
Finally, \system updates the ID-to-disk-location map.  For a moved
vector, the vector ID remains unchanged, while the disk location is
updated.  For a deleted vector, the disk location is marked as a
special tombstone value.  Note that the in-memory IVF lists do not
contain any invalid tombstones, and thus there is no tombstone test
or invalid-record branch in ANNS query processing.

% Because both sets are ordered before pairing, the resulting layout is independent of the order in which deletion requests are supplied.

% Recast snapshots source pages and packed blocks before moving each complete external record and its corresponding RecastQ row.

% The evaluated update workload needs no graph-connectivity repair or global coarse-quantizer/RecastQ retraining.

\Paragraph{Cost Analysis}
For a batch of $N_{ins}$ inserts, \system writes
$O(\tfrac{N_{ins}}{n_B})$ pages to the list files and
$O(\tfrac{8N_{ins}}{B})$ pages to the ID-to-disk-location map.
Therefore, the total amount of disk I/Os is
$O((\tfrac{1}{n_B}+\tfrac{8}{B})N_{ins})$.
Moreover, the computation cost is $O(N_{ins} \cdot nlist \cdot D)$ since the
most computation intensive step, \textsc{AssignLists}, computes the
distances of $nlist$ centroids for each inserted vector.

%For a batch of $N_{del}$ deletes, \system overwrites $N_{del}$ pages
For a batch of $N_{del}$ deletes, \system accesses $2N_{del}$ pages
to the list files and $2N_{del}$ pages in the ID-to-disk-location map
in the worst case, where each deleted vector record resides in its own
page, and $N_{del}$ source entries are moved.  Hence, the total I/O
cost is $O(4N_{del})$.
Moreover, the computation cost is $O(N_{del})$.

\section{Query-Aware Uncertainty-Guided Candidate Selection}
\label{sec:refinement}

After overviewing \system architecture and algorithms, we delve into
the \quic refinement policy in this section.

\begin{definition}[Uncertainty interval] \label{dfn:ui}
Let $d=d(\mathbf{q},\mathbf{x})$ be the true distance between the
query $\mathbf{q}$ and vector $\mathbf{x}$, and $\hat{d}$ be the
estimated distance.  Given a constant $z>0$, the uncertainty interval
of the distance is $[L, U]$, where $L=\hat{d}-z\tau$,
$U=\hat{d}+z\tau$, and $\tau$ is the vector-specific uncertainty scale.
\end{definition}

Our \recastq encoding provides both a deterministic and a
probabilistic uncertainty interval (cf.
Section~\ref{subsec:uncertainty-scale}).  The deterministic interval
is distribution-free, but its bound is quite loose.  The probabilistic
interval reduces the deterministic bound by a factor of $\sqrt{D}$.

Then, \quic selects refinement candidates as follows.

\Paragraph{\quic policy}
Let $S$ be the set of vectors in all probed IVF lists.  Suppose
for $x_i\in S$, the uncertainty interval is $[L_i, U_i]$.  Let
\begin{equation}
\theta_K=K\text{-th-smallest}\{U_i:x_i\in S\}.
\end{equation}
The refinement candidate set is
\begin{equation}
C(S)=\{x_i\in S:L_i\leq\theta_K\}.
\label{eq:selector}
\end{equation}

% The boundary is computed globally across all probed lists. A vector with $L_i>\theta_K$ is worse, even at its lower endpoint, than the upper endpoints of at least $K$ scanned vectors. It therefore cannot affect the exact Top-$K$ when the intervals are valid.

% Section~\ref{subsec:uncertainty-scale} derives RecastQ's estimated distance and uncertainty scale from its directional reconstruction error. This section takes the resulting probabilistic endpoints as inputs and develops the global rule for deciding which scanned vectors still merit exact-vector access. Approximate rank alone gives no query-specific stopping condition.

\subsection{Quality of Refinement Candidates}
\label{subsec:selector}

Within the set $S$ of vectors in the probed IVF lists,
an ideal policy would keep all the true top-$K$ neighbors in $S$
as refinement candidates.  (Note that candidate selection policy
obviously cannot recover a true neighbor missed by IVF list selection.) 
In the following, we would like to understand how well \quic preserves
true neighbors as refinement candidates.

% For nondegenerate endpoints and fixed tie handling, $C(S)$ is also minimal among policies that must preserve Top-$K$ for every interval-consistent distance assignment: any retained interval can be made a Top-$K$ member.  Additional pruning therefore requires tighter uncertainty information, which RecastQ targets directly.

\begin{theorem}[\quic quality under deterministic intervals]
\label{thm:conditional-topk}
If $d_i \in [L_i, U_i]$ for every $x_i\in S$, then $C(S)$ retains all top-$K$ neighbors in $S$.
\end{theorem}

\begin{proof}
For an excluded $\mathbf{x}_i$, $d_i \geq L_i>\theta_K$.  However,
there exists at least $K$ vectors $\mathbf{x}_{j_1}$, ...,
$\mathbf{x}_{j_K}$ s.t. $d_{j_k} \leq U_{j_k} \leq \theta_K$.  Thus,
$\mathbf{x}_i$ cannot be a top-$K$ neighbor.
\end{proof}

For the probabilistic uncertainty interval, we introduce the following
result with a number of assumptions for simplifying the analysis.
(Please see the detailed derivation in Appendix~\ref{app:proofs}.)
First, conditioned on the true distances, the normalized distance-estimation 
errors are \emph{independent} standard Gaussian variables.
Second, the different vectors share a common uncertainty scale $\tau$.
Finally, the upper bounds of all intervals are evenly spaced.  The
space between two subsequent upper bounds is denoted $\Delta$.
 
\begin{theorem}[\quic quality under probabilistic intervals]
\label{thm:integral-recall-bound}
Under the above assumptions, for $z>0$,
\begin{equation}
\mathbb E[R_{\mathrm{ref}}@K] \geq
\exp\!\left( \tfrac{\sqrt{2}\tau}{\Delta} I(\sqrt{2}z) \right).
\label{eq:integral-recall-bound}
\end{equation}
where $I(a)=\int_a^\infty\ln\Phi(x)\ dx$, and $\Phi(x)$ is the CDF of the
standard Gaussian distribution.
\end{theorem}

The theorem provides a lower bound for the expected recall in the
refinement stage (within the set $S$ of vectors in the probed IVF
lists).  
For illustration, for SIFT100M, using the empirical mean ratio
$\frac{\mathbb E[\tau]}{\mathbb E[\Delta]}=20.316$ as a plug-in
approximation to $\frac{\tau}{\Delta}$, Eq.~\ref{eq:integral-recall-bound}
gives $0.9803$ at $z=2$.

While the assumptions are very simplified, this result illustrates the
quality of the probabilistic uncertainty interval to some extent.

\begin{theorem}[Number of refinement candidates] \label{thm:numrefine}
Under the above assumptions, the number of refinement candidates is $K
+ \lfloor \frac{2z\tau}{\Delta} \rfloor$.
\end{theorem}

\begin{proof}
A vector $\mathbf{x_i}$ is selected as a candidate iff. $L_i \leq
\theta_K$.  Under the above assumptions, $U_i=L_i+2z\tau$. Hence, $U_i
\leq \theta_K+2z\tau$.  There are $K$ vectors with upper bounds up to
$\theta_K$, and $\lfloor \frac{2z\tau}{\Delta} \rfloor$ selected vectors, whose upper
bounds in $(\theta_K, \theta_K+2z\tau]$.
\end{proof}

We see that the number of refinement candidates increases with $z$ and
$\tau$, and decreases with $\Delta$.  First, larger $z$ makes the
selection more conservative.  Second, smaller quantization errors lead
to lower $\tau$ and fewer candidates to refine.  Finally, smaller
$\Delta$ means more vectors with similar distances, which makes the
query intrinsically harder and requires refining more candidates.

% This local-spacing interpretation connects the system's three control factors.  A dense Top-$K$ boundary has small $\Delta_U$ and makes a query intrinsically harder; increasing $z$ makes selection more conservative; and reducing directional error lowers $\tau$.  Refinement demand is thus governed by uncertainty relative to local Top-$K$ boundary spacing, rather than by approximate rank alone.  Equation~\ref{eq:refinement-demand} is an explanatory first-order model, not a deterministic guarantee.

\subsection{QUIC Computation}
\label{subsec:quic-compute}

\Paragraph{Computing $\theta_K$}
During the FastScan of packed codes in the probed IVF lists, a
size-$K$ max heap maintains the globally smallest $K$ upper bounds
observed so far.  The heap maximum gives the current $cur\theta_K$,
which is monotonically non-increasing after initialization.  For each
SIMD batch, \system compares all upper bounds against $cur\theta_K$ to
filter out any upper bound that is greater than or equal to
$cur\theta_K$ for reducing heap operations.  The remaining upper
bounds are processed serially: each is rechecked against the latest
heap maximum and, if still eligible, inserted with an $O(\log K)$ heap
update.  The same heap is used across all probed IVF lists, ensuring
that the bound is computed globally.

% This SIMD filtering applies only to boundary maintenance; refinement candidates are selected separately by comparing their lower bounds with the final global $\theta_K$.

\Paragraph{Selecting Refinement Candidates}
After obtaining global $\theta_K$, \system compares the lower bounds
to select the refinement candidates.  To safeguard against unhealthy
growth of per-query workspace, the implementation includes a
2,000-entry transient shortlist of candidates for a query.  It is
almost never activated under the evaluated configurations and has no
measurable effect on the reported Recall--QPS frontiers.

\section{RecastQ Encoding}
\label{sec:encoding}

The \recastq encoding process is as follows:

\begin{list}{}{\setlength{\leftmargin}{5.5mm}\setlength{\itemindent}{-4.0mm}\setlength{\topsep}{0.5mm}\setlength{\itemsep}{0mm}\setlength{\parsep}{0.5mm}}

\item (1) Let $\mathbf{c}$ be a reference center.  (For example, let
$\mathbf{c}$ = $\tfrac{1}{N}\sum_{i=1}^{N} \mathbf{x_i}$, for vectors
$\mathbf{x_1}$, ..., $\mathbf{x_N}$.  $\mathbf{c}$ will not change
upon index updates.) Let $\mathbf{T}$ be an orthogonal transformation.

\item (2) To encode vector $\mathbf{x}$, \recastq computes
$\mathbf{x_c}=\mathbf{x}-\mathbf{c}$, $r = \lVert \mathbf{x_c}
\rVert$, and $\mathbf{u}=\tfrac{\mathbf{x_c}}{r}$, which is the
normalized vector.

\item (3) \recastq encodes $\mathbf{T}\mathbf{u}$ using $M$ code
groups with bit-gain-aware uneven grouping (cf.
Section~\ref{subsec:uneven}). 

\item (4) Three metadata bytes, i.e., \emph{norm}, \emph{scale}, and
\emph{error}, are stored with the main code to facilitate decode
processing.  Let $\mathbf{y'}$ be the decoded vector from the main
code.  $\mathbf{\hat{u}}=\mathbf{T}^\top \tfrac{\mathbf{y'}}{\lVert
\mathbf{y'}\rVert}$ is an estimation of $\mathbf{u}$. Let the residual
$\mathbf{e}=\mathbf{u}-\mathbf{\hat{u}}$.  Then, \emph{norm} records
$r$, \emph{scale} records $\tfrac{r}{\lVert\mathbf{y'}\rVert}$, and
\emph{error} records $\lVert\mathbf{e}\rVert$ (cf.
Section~\ref{subsec:metadata}).

\end{list}
Given a query $\mathbf{q}$, the distance to vector $\mathbf{x}$ is
estimated as follows:
\begin{list}{}{\setlength{\leftmargin}{5.5mm}\setlength{\itemindent}{-4.0mm}\setlength{\topsep}{0.5mm}\setlength{\itemsep}{0mm}\setlength{\parsep}{0.5mm}}

\item (1) Let $\mathbf{q_c}=\mathbf{q}-\mathbf{c}$, and
$\mathbf{q_c'}=\mathbf{T}\mathbf{q_c}$.

\item (2) \recastq obtains the decoded vector $\mathbf{y'}$,
and estimates $r$, $\tfrac{r}{\lVert\mathbf{y'}\rVert}$, and
$\lVert\mathbf{e}\rVert$ from the three metadata bytes. 

\item (3) The squared L2 distance is computed as follows:
\begin{equation}
\hat{d}(\mathbf{q},\mathbf{x})=\lVert\mathbf{q_c}\rVert^2 +r^2
-2\frac{r}{\lVert\mathbf{y'}\rVert}\langle
\mathbf{q_c'},\mathbf{y'}\rangle.
\label{eq:l2-estimate} \end{equation}
We use squared L2 distance in the analysis; other supported distance metrics
are discussed in
\hyperref[par:other-metrics]{Section~\ref*{subsec:uncertainty-scale},
``Other metrics''}.

\item (4) Given a parameter $z>0$, \recastq further computes the
uncertainty interval as
$[\hat{d}(\mathbf{q},\mathbf{x})-z\tau(\mathbf{q},\mathbf{x}) ,
\hat{d}(\mathbf{q},\mathbf{x})+z\tau(\mathbf{q},\mathbf{x})]$, where
$\tau(\mathbf{q},\mathbf{x})=\tfrac{1}{\sqrt{D}}2r\lVert\mathbf{q_c}\rVert
\lVert\mathbf{e}\rVert$ is the uncertainty scale (cf.
Section~\ref{subsec:uncertainty-scale}).

\end{list}

\subsection{Uncertainty Interval}
\label{subsec:uncertainty-scale}

\begin{lemma}[Distance residual]
\label{lem:distance-residual}
The distance estimation satisfies
\begin{equation}
d(\mathbf{q},\mathbf{x})-\hat{d}(\mathbf{q},\mathbf{x})
= -2r\langle\mathbf{q_c},\mathbf{e}\rangle.
\label{eq:exact-residual}
\end{equation}
\end{lemma}

\begin{proof}
We have 
$d(\mathbf{q},\mathbf{x})= 
\lVert \mathbf{q_c}-\mathbf{x_c} \rVert^2 = 
\lVert \mathbf{q_c}-r\mathbf{u} \rVert^2$.  
Since
$\langle \mathbf{q_c'},\tfrac{\mathbf{y'}}{\lVert\mathbf{y'}\rVert} \rangle=
\langle \mathbf{T}\mathbf{q_c},\tfrac{\mathbf{y'}}{\lVert\mathbf{y'}\rVert} \rangle=
\langle\mathbf{q_c},\mathbf{T}^\top\tfrac{\mathbf{y'}}{\lVert\mathbf{y'}\rVert}\rangle=
\langle \mathbf{q_c},\mathbf{\hat{u}} \rangle$,
$\hat{d}(\mathbf{q},\mathbf{x})=\lVert \mathbf{q_c}\rVert^2+r^2
-2r\langle \mathbf{q_c},\mathbf{\hat{u}} \rangle $.
Subtracting the two equations gives Eq.~\ref{eq:exact-residual}.
\end{proof}

Then, the Cauchy--Schwarz inequality immediately gives the
distribution-free bound
$|d(\mathbf{q},\mathbf{x})-\hat{d}(\mathbf{q},\mathbf{x})| \leq
2r\lVert\mathbf{q_c}\rVert \lVert\mathbf{e}\rVert$.
However, this bound assumes the worst-case scenario, and can be too
loose for the selective refinement purpose.  Therefore, we derive
a probabilistic uncertainty bound under the following assumption.

% assumes the worst-case alignment and is $\sqrt D$ times the probabilistic scale used by Recast, making it too wide for selective refinement.

%\begin{equation}
%|d(q,x)-\hat d(q,x)|\leq 2r\lVert q_c\rVert e.
%\label{eq:geometric-envelope}
%\end{equation}

\begin{assumption}[Directional isotropy]
\label{ass:directional-isotropy}

Suppose $r>0$, $\lVert\mathbf{q_c}\rVert>0$, and
$\lVert\mathbf{e}\rVert>0$. The normalized residual
$\tfrac{\mathbf{e}}{\lVert\mathbf{e}\rVert}$ is uniformly distributed
on the $D$-dimensional unit hyper-sphere $\mathbb{S}^{D-1}$.

%relative to the fixed centered-query direction $q_c/\lVert q_c\rVert$.

\end{assumption}

Note that this is a high-dimensional analytical model, not necessarily
a property imposed by the encoding.  If $r=0$,
$\lVert\mathbf{q_c}\rVert=0$, or $\lVert\mathbf{e}\rVert=0$, the
residual in Lemma~\ref{lem:distance-residual} is 0. Then, the
estimated distance is accurate.   

\begin{lemma}[Spherical projection law]
\label{lem:spherical-projection}

Under Assumption~\ref{ass:directional-isotropy}, let $\gamma= \langle
\tfrac{\mathbf{q_c}}{\lVert
\mathbf{q_c}\rVert},\tfrac{\mathbf{e}}{\lVert
\mathbf{e}\rVert}\rangle$.  The probability density function of
$\gamma$ is 
\begin{equation}
f_D(t)= \tfrac{\Gamma(D/2)}{\sqrt{\pi}\,\Gamma((D-1)/2)}
(1-t^2)^{(D-3)/2},\qquad -1<t<1,
\label{eq:spherical-density}
\end{equation}
or equivalently,
$\gamma^2\sim\operatorname{Beta}(1/2,(D-1)/2)$. Moreover,
\begin{equation}
\mathbb{E}(\gamma)=0,\qquad
\operatorname{Var}(\gamma)=\tfrac{1}{D},\qquad
\sqrt{D}\cdot\gamma \xrightarrow{D \to \infty} \mathcal N(0,1).
\label{eq:spherical-limit}
\end{equation}

\end{lemma}

\begin{proof}
$\gamma$ is the projection of $\tfrac{\mathbf{e}}{\lVert
\mathbf{e}\rVert}$ in the direction of $\tfrac{\mathbf{q_c}}{\lVert
\mathbf{q_c}\rVert}$.  Hence, $\gamma$ can be viewed as one coordinate
of a uniform point on $\mathbb S^{D-1}$.  Then,
Eq.~\ref{eq:spherical-density}, its Beta representation, and its
moments follow from the spherical-coordinate
law~\cite{CaiFanJiangAngles}. The normalized coordinate converges to a
standard Gaussian in high
dimensions~\cite{CaiFanJiangAngles,DiaconisFreedmanSphere}.
\end{proof}

\begin{theorem}[RecastQ uncertainty interval]
\label{thm:recast-uncertainty-scale}

Under Assumption~\ref{ass:directional-isotropy},
let
$\tau(\mathbf{q},\mathbf{x})=\tfrac{1}{\sqrt{D}}2r\lVert\mathbf{q_c}\rVert
\lVert\mathbf{e}\rVert$.  Then, as $D$ increases to infinity,
\begin{equation}
\frac{d(\mathbf{q},\mathbf{x})-\hat{d}(\mathbf{q},\mathbf{x})}{\tau(\mathbf{q},\mathbf{x})} 
=-\sqrt{D}\gamma \xrightarrow{D \to \infty} \mathcal N(0,1),
\label{eq:standardized-residual}
\end{equation}
\begin{equation}
\Pr\{|d(\mathbf{q},\mathbf{x})-\hat{d}(\mathbf{q},\mathbf{x})|\leq
z\tau(\mathbf{q},\mathbf{x})\} \xrightarrow{D \to \infty} 2\Phi(z)-1.
\label{eq:marginal-coverage}
\end{equation}

\end{theorem}

\begin{proof}
For Eq.~\ref{eq:standardized-residual},  the left hand side=
$\tfrac{-2r\langle\mathbf{q_c},\mathbf{e}\rangle\sqrt{D}}{2r\lVert\mathbf{q_c}\rVert\lVert\mathbf{e}\rVert}$
= $-\sqrt{D}\gamma$.  Then, Eq.~\ref{eq:standardized-residual} follows
from Lemma~\ref{lem:spherical-projection}, and
Eq.~\ref{eq:marginal-coverage} results from the standard Gaussian
distribution.
\end{proof}

Based on Theorem~\ref{thm:recast-uncertainty-scale}, the true distance
is within the uncertainty interval
$[\hat{d}(\mathbf{q},\mathbf{x})-z\tau(\mathbf{q},\mathbf{x}) ,
\hat{d}(\mathbf{q},\mathbf{x})+z\tau(\mathbf{q},\mathbf{x})]$ with
approximate probability $2\Phi(z)-1$.
The parameter $z$ controls the interval size.  Our experiments in
Table~\ref{tab:interval-diagnostics} empirically test the setting of
$z$ (cf. Section~\ref{subsec:encoding-evaluation}).  The uncertainty
interval covers 68.16--68.71\% cases at $z=1$, and 95.43--95.85\%
cases at $z=2$.  These are close to the 68.27\% and 95.45\%
probabilities in the standard Gaussian\footnote{While confirming the
usefulness of our computed uncertainty interval, the results are not
sufficient to establish the residual isotropy assumption across
candidates.}.

\phantomsection \label{par:other-metrics} \Paragraph{Other metrics}
For inner-product and cosine search, we disable reference centering because
subtracting a reference vector changes both measures. 
We support inner product through negative similarity with uncertainty
scale $\tfrac{1}{\sqrt{D}}r\lVert\mathbf{q}\rVert\lVert\mathbf{e}\rVert$.
For cosine distance ($1-\cos(\mathbf{q},\mathbf{x})$), we use
$\tfrac{1}{\sqrt{D}}\lVert\mathbf{e}\rVert$ as the uncertainty scale.

\subsection{Three Metadata Bytes}
\label{subsec:metadata}

\recastq represents each of \emph{norm}, \emph{scale}, and
\emph{error} in one byte.  For \emph{error}, since it is in $[0,2]$,
\recastq stores $\lceil \tfrac{error}{2}255\rceil$.  For \emph{norm} and
\emph{scale}, 255 is reserved as an overflow sentinel.  If a vector's
\emph{norm} or \emph{scale} falls outside the representable metadata
range, \recastq stores the overflow sentinel.  For such cases, the
uncertainty interval is treated as unbounded, which satisfies the
$\theta_K$ bound, and therefore the corresponding vector record is
always refined.  In this way, we ensure that the metadata saturation
cannot erroneously exclude a vector.  For non-sentinel cases, the
normal distance and interval computation are performed.

%The directional PQ4 code alone does not expose the factors in
%Eqs.~\ref{eq:l2-estimate}--\ref{eq:l2-radius}. RecastQ stores three quantized
%bytes per vector:
%\begin{itemize}
%  \item \textbf{norm}: reconstructs $r=\lVert x-c\rVert$, scaling both the
%  point estimate and uncertainty radius;
%  \item \textbf{scale}: stores $r/\lVert y'\rVert$ and converts the raw PQ
%  reconstruction inner product into the normalized directional term in
%  Eq.~\ref{eq:l2-estimate}; and
%  \item \textbf{error}: estimates $e=\lVert u-\hat u\rVert$, the factor that
%  directly controls Eq.~\ref{eq:l2-radius}.
%\end{itemize}

\subsection{Bit-Gain-Aware Uneven Grouping}
\label{subsec:uneven}

PQ performs even grouping: every dimension group contains the same
number of (e.g., 2) dimensions.  Since a group is encoded with a fixed
number of (e.g., 4) code bits, even grouping treats each dimension
equally in essence.  However, different dimensions can have
significantly different data distributions.  

We propose uneven grouping in \recastq to capture the difference
across dimensions.  The basic idea is to (1) partition the dimensions
(for vectors in the transformed space) into uneven groups, which can
contain different numbers of dimensions, and (2) keep the fixed number
of code bits (i.e., 4 bits) per group.  

The benefits of uneven grouping are as follows.
First, the number of code bits per dimension is lower for groups with
larger number of dimensions.  This reflects the data variance across
dimensions, thereby improving the encoding quality and decreasing the
error $\lVert\mathbf{e}\rVert$.
Second, lower errors result in narrower uncertainty intervals, leading
to smaller number of refinement candidates and better query
throughput.
Finally, the resulting code still follows the fixed 4-bit group size,
which is compatible to Faiss PQ4 FastScan, and amenable to SIMD
acceleration.

% The continuous budgets serve as a grouping signal for constructing $M$ unequal-dimensional sub-quantizers. RecastQ partitions the transformed dimensions so that each group has approximately four bits of aggregate ideal demand, while assigning every group exactly one 4-bit code. This one-nibble-per-group contract makes the resulting layout SIMD-friendly: it preserves a regular nibble stream, 32-row PQ4 packed blocks, and a fixed $M/2$-byte payload. 

Next, we answer the question how to assign dimensions to groups.  Let
us consider the benefit of assigning an extra code bit to a dimension.
Let $\mathcal{D}_j(b)$ denote the expected quantization distortion of
a dimension $j$ under a budget of $b$ code bits. According to the
quantization literature~\cite{GrayNeuhoffQuantization},
$\mathcal{D}_j(b)=\kappa\sigma_j^2 2^{-2b}$, where $\sigma_j^2$ is the
variance of dimension $j$.  Then, the reduction of distortion by
adding an extra bit, which is the marginal gain, is $\mathcal{G}_j(b)
=\mathcal{D}_j(b)-\mathcal{D}_j(b+1) =\tfrac{3}{4}\mathcal{D}_j(b)$.
In the ideal case, if we continuously assign the next bit to the
dimension with the largest marginal gain,  the resulting distortion is
the same across all dimensions.  Suppose the total number of code bits
is $4M$, which is determined by the specified memory budget.  Hence,
we have $b_j=\max(0,\log_2\sigma_j-\lambda)$ and $\sum_j b_j=4M$,
where $\lambda$ is a constant chosen to satisfy the bit budget.  That
is, the number of code bits of a dimension is proportional to the
logarithm of the standard deviation $\log_2\sigma_j$.  For dimensions
whose $\log_2\sigma_j \leq \lambda$, no bit is assigned.

\begin{algorithm}[t]
\caption{Bit-gain-aware uneven grouping.}
\label{alg:uneven-groups}
\begin{algorithmic}[1]
\Require vectors $X\in\mathbb{R}^{N\times D}$, number of 4-bit groups $M$
\State $budget[g]\gets 4$ and $G[g]\gets \emptyset$ for group $g=1$,...$M$
\State compute $\sigma_j$ of $X$ for dimension $j=1$, ..., $D$
\State choose $\lambda$ s.t.
       $b[j]\gets\max(0,\log_2\sigma_j-\lambda)$ and $\sum_j b[j]=4M$
\State sort the dimensions in the descending order of $b[j]$
\For{i = 1 to $D$}
    \State $j \gets $ the dimension with the $i$-th largest demand
    \State $g* \gets \operatorname*{argmax} budget[g]$ 
    \State assign dimension $j$ to group $g*$
    \State $budget[g*] \gets budget[g*] - b[j]$,
           $G[g*]\gets G[g*]\cup \{j\}$
\EndFor
\State \Return $G$
\end{algorithmic}
\end{algorithm}

From the above analysis, $b_j$ can be viewed as the demand of
dimension $j$.  Each of the $M$ code group has a code budget of 4
bits.  We design a greedy algorithm to assign dimensions to groups as
shown in Algorithm~\ref{alg:uneven-groups}.  The algorithm initializes
group budgets and dimension demands (Line 1--3).  Then, it sorts the
dimensions (Line 4) and assigns the dimensions one by one in the
descending order of the demand (Line 5--9).   For each dimension, the
group with the largest remaining budget is chosen to accommodate the
dimension (Line 7), and the group's budget is adjusted accordingly
(Line 9).  Finally, the algorithm returns the group assignment.

\section{Memory-Side Optimizations}
\label{sec:memory-optimizations}

% Challenge 2 requires uncertainty-aware candidate selection to remain cheaper than the external work it eliminates. 

Previously, in Section~\ref{sec:architecture}, we have described how
\system uses the compressed coarse quantizer to accelerate the
distance computation of list centroids for IVF list selection. 
In this section, we focus on the optimization of \recastq code
scanning in the probed IVF lists.

% It then scans RecastQ codes in the probed lists, where the numerical accuracy of the lookup-table (LUT) accumulation directly affects both distance estimation and uncertainty-guided candidate selection.

\Paragraph{Why exact-float accumulation}
Faiss FastScan normally quantizes the LUT entries in uint8. This is
effective for approximate ranking, but introduces an additional
numerical residual.

For a row $m$ ($m=1$, ..., $M$), let $T_m[c_m]$ be the exact float32
entry selected from the LUT, and $\widehat{T_m}[c_m]$ be the
reconstructed value from the uint8 quantization.  Then, the additional
residual is
\begin{equation}
\delta_{\mathrm{LUT}}=\sum_m(\widehat{T_m}[c_m]-T_m[c_m]),\qquad
|\delta_{\mathrm{LUT}}|\leq\sum_m\epsilon_m,
\label{eq:lut-error} \end{equation}
where $\epsilon_m$ bounds the quantization error of row $m$. 
$\delta_{\mathrm{LUT}}$ contributes to $\langle
\mathbf{q_c'},\mathbf{y'}\rangle$ in Eq.~\ref{eq:l2-estimate}.  The
squared-L2 estimate scales this inner product by
$-\tfrac{2r}{\lVert\mathbf{y'}\rVert}$, incurring an additional error
up to $\tfrac{2r}{\lVert\mathbf{y'}\rVert}\sum_m\epsilon_m$.
Ignoring the residual would make the uncertainty interval inconsistent
with the distance, but covering it entails wider intervals and
degraded refinement performance.

To avoid this problem, \system uses the original float32 values in the
LUT.  Our experiments in Figure~\ref{fig:exact-lut}(a) confirms the
practical effect of this design decision: exact-float accumulation
reaches the same recall with fewer refinement candidates compared to
uint8 accumulation (cf.  Section~\ref{subsec:encoding-evaluation}).

\Paragraph{Why SIMD is necessary}
While having the desired numerical accuracy, the scalar exact-float
computation extracts one code and performs one float32 LUT lookup for
every code group of every scanned encoded vector.  
Our experiments in Figure~\ref{fig:exact-lut}(b) shows that scalar LUT
lookup dominates the query time.
Therefore, we develop an exact-float kernel that exploits the AVX-512
SIMD operations to accelerate the exact-float computation.  

% In the following, we develop an exact-float
% kernel that preserves the packed code layout of Faiss FastScan and
% processes all 32 vectors in a block in parallel.

% It therefore gives up the block-level throughput of PQ Fast Scan.

% whereas AVX-512 restores float32 accumulation to the same cost regime as uint8 Fast Scan.

\begin{algorithm}[t]
\caption{AVX-512 exact-float \recastq block FastScan.}
\label{alg:exact-float-pq4}
\begin{algorithmic}[1]
\Require 32-vector packed block $B$; the lookup table in float32, LUT
$T[M][16]$; an even number of groups $M$
\Ensure Unquantized lookup sums of 32 vectors $D[0\ldots31]$
\State $A_{\mathrm{lo}},A_{\mathrm{hi}}\gets\Call{ZeroZmm}{},\Call{ZeroZmm}{}$
\For{$m\gets0,2,\ldots,M-2$}
    \State $p\gets B+32*m/2$
    \State $X_0,X_1\gets\Call{Vmovdqu}{p},\Call{Vmovdqu}{p+16}$
    \State $(Q_0^{\mathrm{lo}},Q_1^{\mathrm{lo}})\gets
        \Call{Vpand}{(X_0,X_1),\mathtt{0x0f}}$
    \State $(H_0,H_1)\gets\Call{Vpsrlw}{(X_0,X_1),4}$
    \State $(Q_0^{\mathrm{hi}},Q_1^{\mathrm{hi}})\gets
        \Call{Vpand}{(H_0,H_1),\mathtt{0x0f}}$
    \State $L_0,L_1\gets\Call{Vmovups}{T[m,:]},\Call{Vmovups}{T[m+1,:]}$
    \State $A_{\mathrm{lo}}\gets\Call{LookupAdd}{A_{\mathrm{lo}},Q_0^{\mathrm{lo}},L_0}$
    \State $A_{\mathrm{lo}}\gets\Call{LookupAdd}{A_{\mathrm{lo}},Q_1^{\mathrm{lo}},L_1}$
    \State $A_{\mathrm{hi}}\gets\Call{LookupAdd}{A_{\mathrm{hi}},Q_0^{\mathrm{hi}},L_0}$
    \State $A_{\mathrm{hi}}\gets\Call{LookupAdd}{A_{\mathrm{hi}},Q_1^{\mathrm{hi}},L_1}$
\EndFor
\State $\pi\gets[0,2,4,\ldots,14,1,3,5,\ldots,15]$
\State $D[0\ldots15]\gets\Call{Vpermps}{\pi,A_{\mathrm{lo}}}$
\State $D[16\ldots31]\gets\Call{Vpermps}{\pi,A_{\mathrm{hi}}}$
\State \Return $D$
\end{algorithmic}
\end{algorithm}

\Paragraph{AVX-512 exact-float FastScan}
For the packed \recastq code block, we follow the layout of the Faiss
PQ4 FastScan.  Specifically, the layout requires the number of code
groups $M$ to be an even number.  Then, for every pair of $(m,m+1)$
code groups, the packed block contributes 32 bytes, storing 4-bit code
words for 32 vectors in the two code groups. The first and second 16
bytes encode subquantizers $m$ and $m+1$, respectively.  Within each
byte, the lower 4-bit nibble belongs to vectors 0--15 and the higher
4-bit nibble to vectors 16--31.  Within either 16-vector half, the
lanes are stored in the order of $[0,8,1,9,\ldots,7,15]$.

For the float32-based LUT, an LUT row stores 16 values, which
correspond to the 16 centroids of a 4-bit code group.  Hence, a row
takes 64B, which fits into a 512-bit ZMM register.  Then, inside \texttt{LookupAdd}, 
we use the SIMD permutation \texttt{D=VPERMPS(I, L)} to perform an entire 16-way
lookup inside the ZMM register $L$, where $D[j]$ receives $L[I[j]]$.
For example, if $L$ contains the $T_m$ row in the LUT, indices
$[3,10,1,\ldots]$ obtain $[T_m[3],T_m[10],T_m[1],\ldots]$.

Algorithm~\ref{alg:exact-float-pq4} lists the SIMD-accelerated
exact-float FastScan procedure.  It first initializes two 512-bit ZMM
registers (i.e., $A_{\mathrm{lo}}$ and $A_{\mathrm{hi}}$) as the
accumulators (Line 1), then enters the main for-loop to process $M$
code groups (Line 2--12).
Each for-loop iteration processes a pair of $(m,m+1)$ code groups.  It
obtains the starting address (Line 3), and loads the first and second
16 bytes into two 128-bit XMM registers (i.e., $X_0$ and $X_1$)
corresponding to group $m$ and $m+1$, respectively (Line 4). After
that, the procedure uses bit-wise AND and right-shift operations to
obtain the lower and higher nibbles (Line 5--7), loads the two LUT
rows (Line 8), and looks up and accumulates the LUT values to
$A_{\mathrm{lo}}$ and $A_{\mathrm{hi}}$ (Line 9--12).  Here,
$LookupAdd(A,Q,L)$ widens the 16 bytes in $Q$ to 16 32-bit indices
with $VPMOVZXBD$, performs 16 register lookups with $VPERMPS$, and
accumulates the retrieved values with $VADDPS$.
Finally, the procedure shuffles and returns the decoded values for the
32 vectors in the packed block (Line 13--16).

\begin{figure*}
  \centering
  \includegraphics[width=0.8\textwidth]{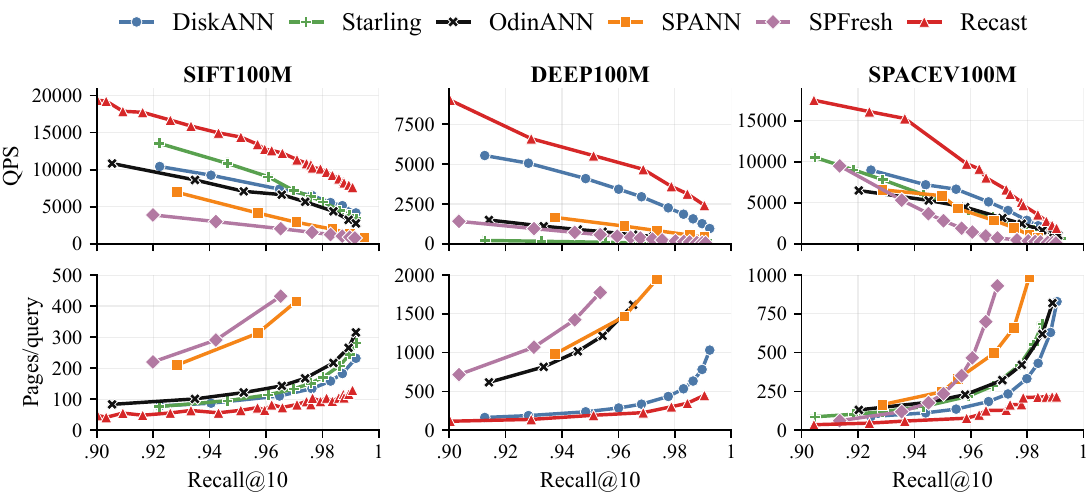}
  \vspace{-0.15in}
  \caption{Static top-10 query performance.}
  \label{fig:static-query}
  \vspace{-0.15in}
\end{figure*}

\Paragraph{Metadata bytes and SIMD-based \quic computation}
For each 32-vector code block, \recastq stores three 32-byte arrays
for the encoded \emph{norm}, \emph{scale}, and \emph{error}.  
During the \recastq FastScan, the metadata bytes are processed 16
vectors at a time.  For each 16 vectors, a 16-byte array is loaded
into a 128-bit XMM register. Then, each 8-bit code is widened to 32
bits and decoded in a 512-bit ZMM register.  After that, the estimated
distance $\hat{d}$ and uncertainty interval $[L, U]$ are computed in
SIMD.

Given the computed $[L, U]$, \quic uses a masked SIMD comparison to
reject upper bounds above $cur\theta_K$.  The remaining upper bounds
are inserted to the heap.
\quic performs a second SIMD comparison to filter vectors with lower
bounds above $cur\theta_K$.  For the surviving candidate vectors,
\quic appends their estimated distances, lower and upper bounds, and
computed in-list offsets to a buffer.
After the \recastq FastScan, \quic obtains the global $\theta_K$, and
re-compares the lower bounds of vectors in the buffer against
$\theta_K$ to obtain the final refinement candidates.

% \texttt{VCOMPRESSPS} and \texttt{VPCOMPRESSD} append only surviving estimates, uncertainties, and IDs to aligned SoA arrays, while lower bounds are evaluated only for the final top-$R$ reservoir.  This fused path avoids materialized per-vector interval records and per-lane metadata gathers, and moves most admission decisions to SIMD masks.

In summary, the SIMD-accelerated procedure fuses metadata decoding,
reconstruction of $\hat{d}$ and $[L, U]$, global-bound update, and
refinement candidate selection, while avoiding the materialization
cost of per-vector uncertainty intervals for most vectors.

\section{Evaluation}
\label{sec:evaluation}

% Archived detailed RQ wording; superseded by the concise scope below.
\begin{comment}
We answer five questions: (RQ1) can \system retain competitive high-recall
serving performance; (RQ2) can it process large replacement batches while
preserving post-update behavior; (RQ3) why does it require few exact-vector
accesses; (RQ4) what construction cost, query-ready memory, and external
footprint follow from its organization; and (RQ5) how do serving and
maintenance throughput scale?
\end{comment}

In this section, we perform extensive experiments to compare \system
with state-of-the-art storage-optimized ANNS systems.  We first
describe the experimental setup in Section~\ref{subsec:setup}.  Then,
we study the static query performance and the index performance under
dynamic updates in
Section~\ref{subsec:end-to-end}--\ref{subsec:update}, and the
effectiveness of refinement, storage footprint, and scalability in
Section~\ref{subsec:encoding-evaluation}--\ref{subsec:scalability}.

\subsection{Experimental Setup}
\label{subsec:setup}

\Paragraph{Hardware and common protocol}
The server is equipped with two 36-core Intel Xeon Platinum 8360Y
processors (2.4GHz, 3.5GHz Turbo), 1 TiB DDR4 memory, and Intel NVMe
SSDs.  It runs CentOS 7.9 and supports AVX-512.  All systems are
compiled with GCC 9.3.1 and \texttt{-O3} flag.  Unless stated
otherwise, all systems run 10,000 queries with 72 workers pinned to
distinct physical cores.  Each worker issues 4-KiB-aligned I/Os to the
same SSD with the \texttt{O\_DIRECT} setting and QD=8.  We fix the CPU
frequency at 2.4 GHz and monitor \texttt{Bzy\_MHz}; no AVX-512
frequency reduction is observed.

\Paragraph{Data sets and metrics}
Storage experiments use the 100M variants of SIFT, DEEP, and
SPACEV~\cite{Sift_Gist,Deep1B,SPACEV1B}, whose dimensionalities are
128, 96, and 100, respectively; encoding experiments use
their 1M variants. The distance metric is squared L2 in all results.
We report Recall@$K$, QPS, refined vectors, the number of read
requests, and/or unique 4-KiB storage pages as performance metrics.
Fixed-recall values are interpolated from adjacent measured Pareto
points; there is no extrapolation.

\begin{figure*}
  \centering
  \includegraphics[width=0.8\textwidth]{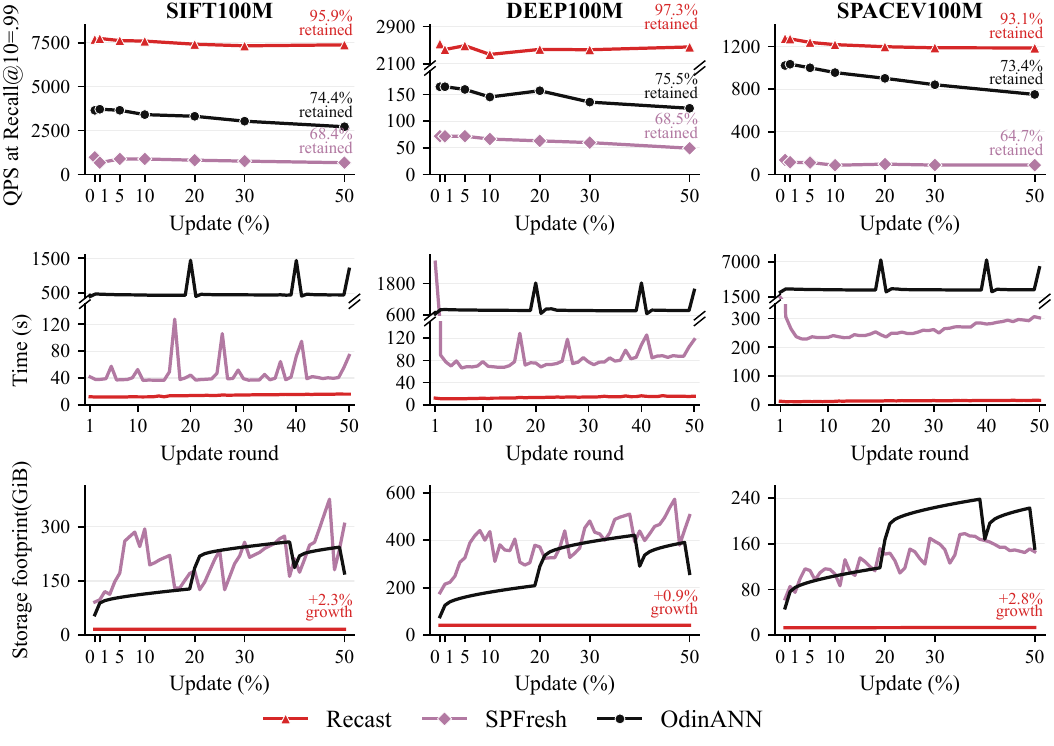}
  \vspace{-0.2in}
  \caption{Query performance, update performance, and storage
footprint with updates.}
  \label{fig:maintenance-updates}
  \vspace{-0.15in}
\end{figure*}

\Paragraph{Systems and fairness}
We compare \system with DiskANN~\cite{DiskANN},
Starling~\cite{Starling}, OdinANN~\cite{OdinANN}, SPANN~\cite{SPANN},
and SPFresh~\cite{SPFresh}.   For each system, we tune its parameters
to optimize the query performance.
DiskANN uses $R=64$, build $L=100$, no node cache, and beamwidth 8.
Starling uses page search with $use\_ratio$=0.3 and beamwidth
8. Its disk-graph $(R,L)$, and PQ
bytes/vector are $(64,128,19)$, $(31,128,8)$, and
$(64,128,12)$ on SIFT100M, DEEP100M, and SPACEV100M,
respectively. Its navigation graph uses
$(R_m,L_m,\alpha)=(32,128,1.2)$; the navigation sample rates are
0.25\%, 0.1\%, and 0.1\%.
Among the settings, $L$, $L_m$, and
$\alpha$ follow the public configuration, $use\_ratio$ and the 8
passes follow the paper's recommended settings, while we tune the rest
of the settings.
SPANN uses ratios 0.10, 0.026, and 0.09 on SIFT100M,
DEEP100M, and SPACEV100M, respectively.
SPFresh uses ratios 0.065, 0.020, and 0.080 on the three data sets,
respectively, with a 256-MiB RocksDB cache.
OdinANN uses the official PipeANN 0.1.1 implementation with $R=96$,
build $L=128$, no in-memory graph, and pipeline width 8. Its
$(B,\text{PQ bytes/vector})$ settings are $(2.25,21)$ on SIFT100M
and $(1.785,16)$ on both DEEP100M and SPACEV100M. 
\system uses 100K IVF lists, with \recastq $M=60$, $M=44$, and
$M=44$ on SIFT100M, DEEP100M, and SPACEV100M, respectively.

% Detailed memory-accounting definition retained for reproducibility:
% the D/40 budget applies to L-B, where B is process RSS immediately before
% native index loading and L is RSS immediately after it. Serving additionally
% requires query buffers, worker-local state, I/O contexts, and configured
% caches. The query-ready footprint reported in Table~\ref{tab:construction}
% is W-B, where W is measured after warm-up and worker initialization but
% before the timed queries. P-B denotes the formal-query peak when collected.

Moreover, for each 100M data set of dimension $D$, we set the memory
budget to be $D/40$ GiB (3.2 GiB for SIFT100M, 2.4 GiB for DEEP100M,
and 2.5 GiB for SPACEV100M).  Both in-memory index and configured
caches count toward this budget.  The memory footprint is measured as
the increase of the process's memory before and after loading the
index.  To reduce the configuration effort, we allow some compared 
systems to exceed this budget.  For DEEP, Starling and
SPANN each exceed the budget by $\sim$63 MiB. For SPACEV, SPANN and
SPFresh exceed the budget by 824 and 336 MiB, respectively.

Furthermore, we compare \recastq with three representative
quantization encodings, i.e., PQ, OPQ, and RabitQ.  PQ and OPQ use the
Faiss implementation~\cite{Faiss}. RaBitQ uses
RaBitQ-Library~\cite{RaBitQLibrary} and its native reranking rule in
the papers~\cite{RabitQ,ExtendedRabitQ}.

\subsection{Static Query Performance}
\label{subsec:end-to-end}

Figure~\ref{fig:static-query} reports top-10 static query performance.
Each column in the figure corresponds to a data set.  The upper row
reports the measured Recall--QPS frontier, while the lower row reports
the number of 4-KiB disk page accesses per query.

We see that \system achieves the best performance across all data
sets.  At Recall@10=0.99, \system achieves 1.56--2.01$\times$ the QPS
of the strongest baseline with the same memory budget, while reading
1.73--4.10$\times$ fewer pages.  This suggests that the high-recall
throughput of \system comes from the reduction of the number of I/Os.
Moreover, as the recall increases, the number of I/Os of \system grows
only slightly.  Higher recalls typically require larger
\texttt{nprobe}, i.e., probing more IVF lists, while our \recastq and
\quic policy effectively filter out unnecessary exact-vector
refinement.

Posting list based solutions incur much higher number of disk I/Os at
high recalls because of retrieving whole selected postings, resulting
in less competitive query performance.
In contrast, graph-based solutions issue fewer I/Os and are generally
more competitive than posting list based systems in the experiments.

Interestingly, DiskANN out-performs Starling in some cases.  Compared
to DiskANN, Starling allocates part of its memory budget to the
navigation graph and the mapping structures.  As a result, the PQ code
length is reduced.  This pressure is the strongest on DEEP100M because
DEEP100M contains float32 vectors compared to the uint8 vectors in
SIFT100M and SPACEV100M.  Starling has to use 8B PQ codes, incurring
significant accuracy degradation of the PQ-based distance estimation.
(Note that this is a memory-budget-dependent trade-off, not an
inherent claim that Starling is weaker than DiskANN.)

\subsection{Maintenance Efficiency and Update Stability}
\label{subsec:update}

We start from the static 100M indexes, and perform 50 rounds of
updates.  Each round deletes 1M vectors and inserts 1M new vectors,
preserving a total 100M active vectors.  As shown in
Figure~\ref{fig:maintenance-updates}, after cumulative updates have
replaced 0\%, 1\%, 5\%, 10\%, 20\%, 30\%, and 50\% of the corpus, we
report the top-10 query throughput at recall=0.99 in the top row, and
the storage footprint in the bottom.  For the update performance, we
report the time for completing the each 1M delete+1M insert round in
the middle row.

From the figure, we see that \system retains 93.1--97.3\% of its
round-0 QPS after replacing half of each measured corpus. Its update
rounds take an average 13.6--14.0 seconds, while its storage footprint
grows by only 0.9--2.8\%. 
Moreover, compared to SPFresh, \system lowers the average update latency by
3.3--20.7$\times$; compared to OdinANN, it lowers the average by
36.1--210.5$\times$.  OdinANN retains 73.4--75.5\%
of its initial QPS after 50\% corpus replacement, 
and SPFresh retains 64.7--68.5\%.
Furthermore, there are non-monotonic points and latency spikes in the
results of SPFresh and OdinANN.  They often coincide with the
compaction or merging activities.  For OdinANN, its
active-plus-standby storage work set grows by 204.9--234.2\% across the three datasets and
oscillates as merged versions replace old ones.  In comparison,
\system shows robust performance because \system does not perform
background tasks for index maintenance.

\subsection{Refinement Effectiveness}
\label{subsec:encoding-evaluation}

Figure~\ref{fig:encoding-refinement} studies the effectiveness of the
encoding methods and refinement policies.  In the left sub-figures, we
compare \recastq with PQ, OPQ, and RabitQ. Both PQ and OPQ use the
fixed-$R$ policy.  In the right sub-figures, we use \quic to enhance
both PQ and OPQ, and compare the baseline PQ/OPQ and the enhanced
PQ/OPQ.
All methods probe the same IVF lists. We set \texttt{nprobe} to
62/78/245 for SIFT/DEEP/SPACEV, giving unencoded coarse Recall@10 of
0.99524/0.99500/0.99007.  The vertical dashed line in each sub-figure
marks this ideal recall ceiling.

\begin{figure}
  \centering
  \includegraphics[width=\columnwidth]{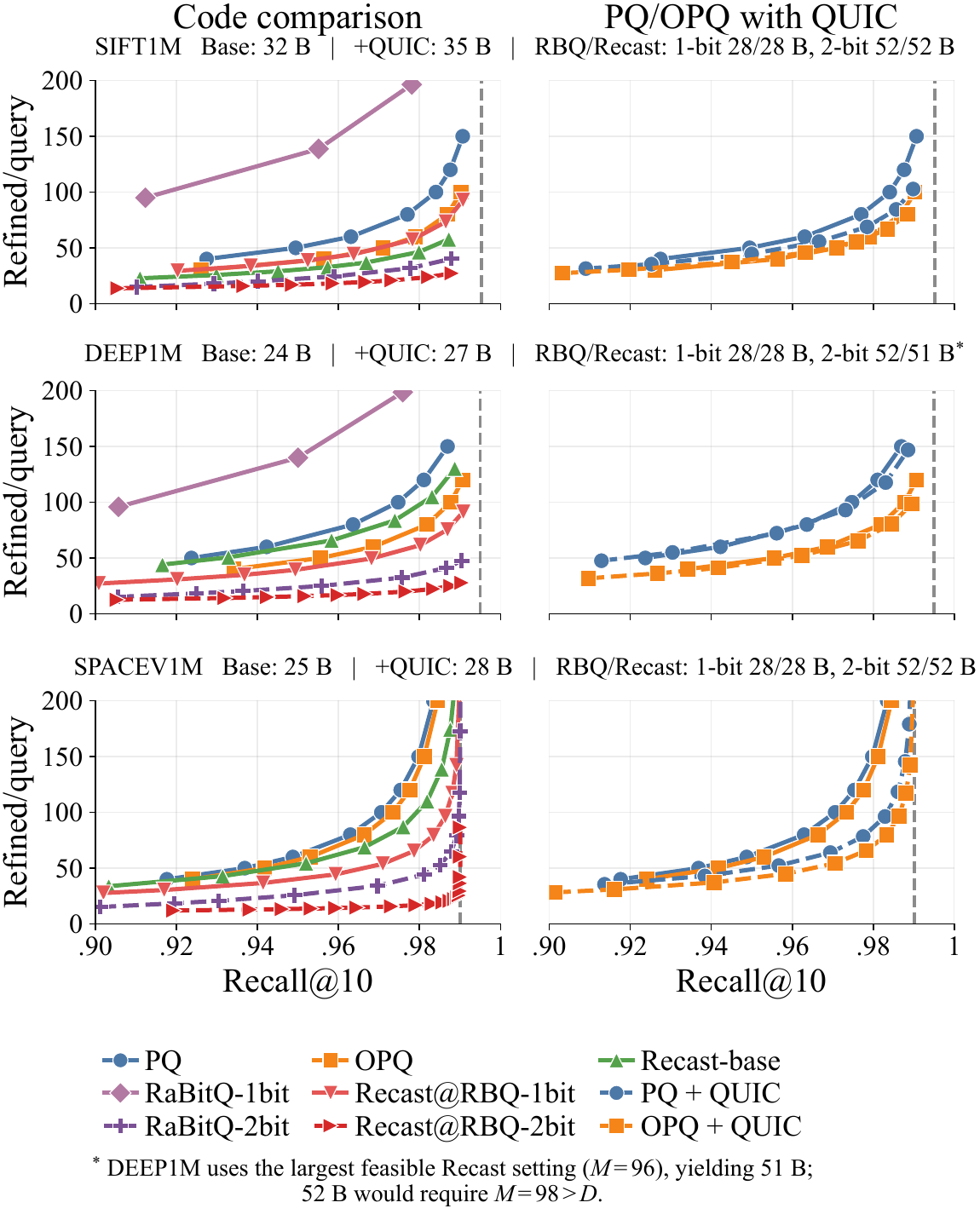}
  \vspace{-0.30in}
  \caption{Effectiveness of \recastq and \quic policy.}
  \label{fig:encoding-refinement}
  \vspace{-0.20in}
\end{figure}

From the left sub-figures, we see that \recastq is the overall best
encoding method across all data sets.  At matched encoding sizes,
\recastq refines 3.4--7.9$\times$ and 1.5--2.5$\times$ fewer number of
vectors than 1-bit RaBitQ at 28 bytes/vector and 2-bit RaBitQ near 52
bytes/vector, respectively.  The 2-bit RaBitQ is the best performing
encoding among the baseline methods.  RaBitQ's reported sizes include
its required dimension padding and auxiliary factors.  For SPACEV1M,
RaBitQ-1bit is absent because all of its measured points at
Recall@10$\geq$0.90 exceed the Y-axis limit of 200 refinements/query.

From the right sub-figures, we see that applying \quic to PQ/OPQ leads
to improved performance.  (Solid curves are the baseline methods,
while dashed curves correspond to the enhanced methods.) The
improvement is especially pronounced for SPACEV.  This confirms the
benefit of \quic compared to the fixed-$R$ policy.

% separate view of uncertainty-guided selection versus directional-code design; its benefit depends on how informative the compact metadata remain for each encoding.  

% DEEP is the main narrow-budget exception: its vectors are already unit-normalized, so norm and radial-scale metadata provide little additional candidate information. At 24 bytes/vector, OPQ can therefore devote the full budget to directional codes while \recastq reserves three bytes for largely redundant metadata. Our claim is a stronger refinement-oriented frontier at practical matched budgets, not universal quantizer dominance.

% Archived point-by-point encoding-size narration; superseded by the aggregate
% memory--refinement result below.
\begin{comment}
Figure~\ref{fig:size-refine} fixes $n_{\mathrm{list}}=1024$ and
$n_{\mathrm{probe}}=64$, then varies only \recastq's encoding size and interval
parameter. Each plotted value is linearly interpolated between two measured
Recall--refinement points; no point is extrapolated. Larger codes reduce
refinement at every target, with a stronger effect at high recall. For
readability, the plot omits points above 200 and 600 refinements/query on SIFT
and DEEP, respectively. Within the displayed Recall@10=.99 curves, increasing
the encoding from 21 to 35 bytes reduces SIFT1M's demand from 178.2 to 53.2
vectors/query, while increasing it from 17 to 27 bytes reduces DEEP1M's demand
from 496.4 to 116.9. The curves flatten toward larger codes, exposing a direct,
workload-dependent memory--I/O tradeoff.
\end{comment}

Figure~\ref{fig:size-refine} shows the impact of encoding size on the
number of refined vectors for \recastq at different recalls.  We see
that as the code size increases, the number of refined vectors quickly
decreases.  At Recall@10=0.99, the plotted code-size ranges reduce
refinement demand by 3.3$\times$ on SIFT1M and 4.2$\times$ on DEEP1M.
This is because larger code leads to more accurate estimation of the
distance and the uncertainty intervals.

\begin{figure}[t]
  \centering
  \includegraphics[width=\columnwidth]{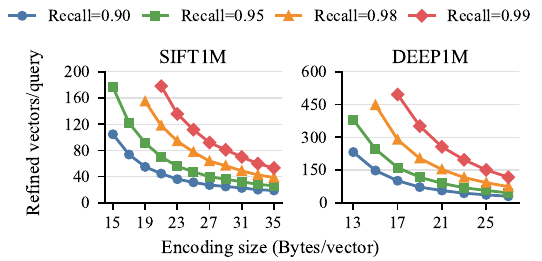}
  \vspace{-0.35in}
  \caption{Impact of code size on refinement performance.}
  \label{fig:size-refine}
  \vspace{-0.15in}
\end{figure}

% Archived stage-by-stage LUT narration; superseded by the concise ablation
% result below.
\begin{comment}
\system's AVX-512 path accumulates an unquantized float32 lookup table over the
same packed PQ4 codes. Here ``exact'' modifies LUT accumulation, not the final
vector distance. Figure~\ref{fig:exact-lut} uses the SIFT100M $M=60$ index.
The left panel fixes $n_{\mathrm{probe}}=421$ and sweeps $z$ under AVX-512
float32 and uint8 LUT accumulation; exact float32 traces the better
Recall--refinement frontier. Points above 180 refinements/query are omitted.
At the Recall@10=.99 operating point, the right panel decomposes single-worker
latency into list selection, Fast Scan,
\quic, and I/O (QD=8). AVX-512 takes 6.0 ms/query, versus 6.8 ms for uint8 and
134.5 ms for scalar float32. Scalar lookup alone costs 130.7 ms, whereas the
AVX-512 Fast Scan costs 2.39 ms. At QD=8, AVX-512's memory-side stages total
4.4 ms/query, exceeding the 1.6-ms I/O stage; optimizing resident computation
is therefore necessary to expose storage parallelism. SIMD preserves the
exact-float score while removing its otherwise prohibitive execution cost.
\end{comment}

\Paragraph{Float32 vs. uint8 accumulation}
Figure~\ref{fig:exact-lut-recall} compares the refinement performance
for uint8 and float32 accumulation designs.  We see that that the
float32 accumulation achieves a stronger Recall--refinement frontier
compared to uint8 accumulation.  This motivates us to use float32
values in the LUT.

Figure~\ref{fig:exact-lut-time} reports the breakdown of the execution
cost at Recall@10=0.99.  For the scalar float32 design, \recastq scan
dominates the query time.  Hence, we exploit SIMD acceleration to
reduce the computation cost.  The resulting AVX-512 based design takes
6.0 ms/query, compared with 6.8 ms for uint8 and 134.5 ms for scalar
float32, a 22.4$\times$ speedup over the scalar path.  Interestingly,
the uint8 SIMD design in Faiss takes slightly longer time than our
float32 SIMD design.  This is because the uint8 SIMD design needs to
perform an extra decoding step for the uint8 codes in LUT.

% With QD=8, the in-memory stages still take 4.4 ms/query versus 1.6 ms for I/O, confirming that efficient FastScan and \quic execution are necessary to expose the storage parallelism.

\Paragraph{\quic with and without SIMD}
Figure~\ref{fig:quic-simd-ablation} uses the same SIFT100M
single-worker, QD=8 operating point as
Figure~\ref{fig:exact-lut-time}.  The time is also broken down in the
same fashion as Figure~\ref{fig:exact-lut-time}.  

From Figure~\ref{fig:quic-simd-ablation}, we see that enabling \quic's
SIMD pipeline reduces the green \quic stage from 6.42 to 1.42 ms/query
(i.e., a 4.52$\times$ improvement) and the total time from 11.86 to
5.96 ms/query (i.e., a 1.99$\times$ improvement).  In this experiment,
the recall, refinement candidate count, and the number of I/O page
accesses remain unchanged.  Therefore, the performance gain clearly
comes from the accelerated in-memory execution.

\begin{figure}[t]
  \centering
  \begin{subfigure}[t]{0.485\columnwidth}
    \centering
    \includegraphics[width=\linewidth]{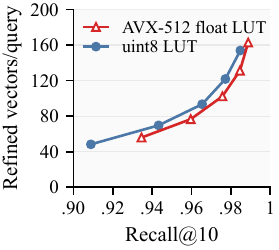}
    \vspace{-0.20in}
    \caption{Recall--refinement frontier.}
    \label{fig:exact-lut-recall}
  \end{subfigure}\hfill
  \begin{subfigure}[t]{0.485\columnwidth}
    \centering
    \includegraphics[width=\linewidth]{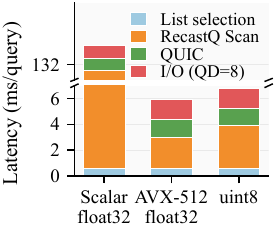}
    \vspace{-0.20in}
    \caption{Query-time breakdown.}
    \label{fig:exact-lut-time}
  \end{subfigure}
  \vspace{-0.15in}
  \caption{LUT accuracy and execution cost.}
  \label{fig:exact-lut}
  \vspace{-0.10in}
\end{figure}

\begin{table}[t]
\caption{Interval calibration and metadata overflow.}
\label{tab:interval-diagnostics}
\vspace{-0.15in}
\centering
%\small
\begin{tabular}{lrrr}
\toprule
\textbf{Data set} &
\shortstack{\textbf{Coverage}\\\textbf{$z=1$ (\%)}} &
\shortstack{\textbf{Coverage}\\\textbf{$z=2$ (\%)}} &
\shortstack{\textbf{Metadata}\\\textbf{overflow (\%)}} \\
\midrule
SIFT1M & 68.16 & 95.43 & 0.0012 \\
DEEP1M & 68.71 & 95.85 & 0.0004 \\
SPACEV1M & 68.37 & 95.67 & 0.0001 \\
\bottomrule
\end{tabular}
\vspace{-0.20in}
\end{table}

% Archived diagnostic setup and nominal-reference narration; superseded by the
% concise calibration result below.
\begin{comment}
Table~\ref{tab:interval-diagnostics} evaluates the deployed uncertainty scale
over all scanned query--vector pairs using production-decoded metadata and the
same $n_{\mathrm{probe}}$ values as Figure~\ref{fig:encoding-refinement}. We set
$M=D/2-6$, making the complete $M/2+3$-byte \recastq encoding exactly $D/4$
bytes/vector. Under the high-dimensional model in
Section~\ref{subsec:uncertainty-scale}, the normalized residual
$r=(d-\hat d)/\tau$ is approximately standard normal, whose nominal two-sided
coverage is 68.27\% at $z=1$ and 95.45\% at $z=2$. Measured Coverage@$z=1$
ranges from 68.16\% to 68.71\%, while Coverage@$z=2$ ranges from 95.43\% to
95.85\%. Thus, the deployed intervals retain coverage close to their Gaussian
nominal references after metadata quantization. Metadata overflow is below
0.002\% throughout, so the conservative overflow path is rarely exercised.
These per-pair measurements diagnose the uncertainty model rather than
Recall@10.
\end{comment}

\Paragraph{Quality of uncertainty interval}
Table~\ref{tab:interval-diagnostics} verifies the coverage of the
uncertainty intervals for the three data sets.  We see that the
uncertainty intervals cover 68.16--68.71\% and 95.43--95.85\% of the
true distances at $z=1$ and $z=2$, respectively.  This is quite close
to the Gaussian based estimation, which is 68.27\% and 95.45\% at
$z=1$ and $z=2$, respectively\footnote{ Note that these results
empirically corroborate the probabilistic uncertainty intervals under
Assumption~\ref{ass:directional-isotropy}.  However, they are not
sufficient to confirm the assumption.  Nor do they test independence
across candidates.  }.

Table~\ref{tab:interval-diagnostics} also shows the percentage of
vectors with overflowed \emph{norm} or \emph{scale}.  We see that
metadata overflow remains below 0.002\%, showing that quantized
uncertainty information rarely invokes the conservative overflow path.

\subsection{Storage Footprint}
\label{subsec:construction}

% Archived component-level storage accounting; superseded by the concise scope
% statement below.
\begin{comment}
Table~\ref{tab:construction} reports the storage footprint required by
each query-ready index. For graph systems, this includes the storage-resident
graph; for posting-list systems, it includes full postings or RocksDB state; for
\system, it includes list-local logical-ID/exact-vector records and the update
directory. Resident PQ codes, head indexes, partitions, and navigation sidecars
are excluded because they are already covered by the loaded-index memory
budget.
\end{comment}

Table~\ref{tab:construction} reports query-required storage data,
excluding structures already counted in the common in-memory index
budget.  We see that \system has the lowest storage footprint for
every data set, using 1.22--2.88$\times$ less storage space than the
system with the next lowest storage footprint. 
(We construct the 100M-vector indexes of \system in 14.1--21.5 minutes
using multiple GPUs to perform the K-means clustering and parallel CPU
threads to produce the IVF lists.  Here, the build time is mentioned
to provide context for understanding \system rather than for an
equal-hardware comparison.)

\begin{table}[t]
\caption{Storage footprint on 100M-vector data sets (GiB).}
\label{tab:construction}
\vspace{-0.15in}
\centering
\renewcommand{\arraystretch}{1.06}
\begin{tabular*}{\columnwidth}{@{\extracolsep{\fill}}lrrr@{}}
\toprule
\textbf{System} & \textbf{SIFT100M} & \textbf{DEEP100M} & \textbf{SPACEV100M} \\
\midrule
DiskANN  & 38.147 & 63.578 & 34.679 \\
Starling & 38.147 & 47.684 & 36.597 \\
OdinANN  & 54.496 & 76.294 & 47.684 \\
SPANN    & 85.717 & 234.613 & 60.187 \\
SPFresh  & 89.277 & 178.553 & 63.696 \\
\system   & \textbf{13.236} & \textbf{39.064} & \textbf{13.083} \\
\bottomrule
\end{tabular*}
\vspace{-0.15in}
\end{table}

% Archived per-data set footprint and build narration; superseded by the
% aggregate result below.
\begin{comment}
\system has the smallest external footprint on every data set. Relative to the
next-smallest measured system, it uses 2.88$\times$, 1.22$\times$, and
2.65$\times$ less storage on SIFT100M, DEEP100M, and SPACEV100M, respectively.
This reduction follows from avoiding an item-level external graph and padded or
replicated posting state: external storage contains only the exact records and
their compact update directory. For context, \system constructs the three
100M-vector indexes in 18.0, 21.5, and 14.1 minutes, respectively, combining
multi-GPU k-means with 128-thread CPU assignment, encoding, and materialization;
we report these heterogeneous build times descriptively rather than as an
equal-hardware speedup.
\end{comment}

\subsection{Query and Update Scalability}
\label{subsec:scalability}

% Archived worker-by-worker scalability narration; superseded by the aggregate
% scaling result below.
\begin{comment}
Figure~\ref{fig:static-scalability} separates moderate- and high-recall static
query scaling. Each $T$-worker process is restricted to $T$ distinct physical
cores, with one logical CPU per core. For each system and recall target, we
reuse the same two validation-selected bracketing configurations at every
worker count and linearly interpolate QPS; all runs use QD=8, \texttt{O\_DIRECT},
and \system's primary $M=60$ index. To make the two evaluations directly
reproducible, each 72-worker endpoint reuses the corresponding fixed-recall
interpolation from Figure~\ref{fig:static-query}, rather than an independent
rerun. At Recall@10=.95, \system scales from 293 to 14,464 QPS
(49.3$\times$); graph traversal remains faster at low concurrency, while
\system leads at 72 workers by 1.39$\times$ over Starling and 1.70$\times$ over
DiskANN. At Recall@10=.99, \system scales from 171 to 7,801 QPS
(45.6$\times$), becomes the throughput leader at 32 workers, and exceeds
DiskANN, Starling, SPANN, and SPFresh at 72 workers by 1.71$\times$,
2.03$\times$, 5.71$\times$, and 8.72$\times$, respectively. Thus \system does
not dominate low-concurrency operation, but its in-memory selection and
independent exact reads scale especially well at high recall and concurrency.
\end{comment}

We study the scalability of the storage-optimized ANNS solutions in
this subsection.  We vary the number of query worker threads from 1 to
72.  Figure~\ref{fig:static-scalability} shows the QPS at recall=0.95
and at recall=0.99 while varying the number of query workers.  We use
the same fixed-recall configurations and physical-core protocol at
every worker count. 
We see that as the number of workers increases from 1 to 72, \system
scales QPS by 49.3$\times$ at Recall@10=.95 and 45.6$\times$ at .99.
At 72 workers it exceeds the strongest baseline by 1.39--1.71$\times$
across the two recall targets. Graph traversal remains competitive at
low concurrency, while \system's in-memory selection and independent
exact-vector reads provide the stronger high-concurrency scaling.

\begin{figure}[t]
  \centering
  \includegraphics[width=0.9\columnwidth]{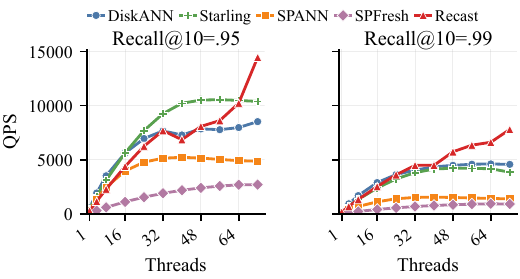}
  \vspace{-0.15in}
  \caption{Static query scalability on SIFT100M.}
  \label{fig:static-scalability}
  \vspace{-0.20in}
\end{figure}

\begin{figure}[t]
  \centering
  \begin{minipage}{0.48\columnwidth}
    \centering
    \includegraphics[width=\linewidth]{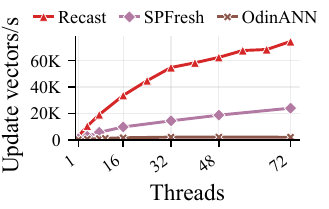}
    \vspace{-0.3in}
    \caption{Update scalability (SIFT100M).}
    \label{fig:update-scalability-panel}
  \end{minipage}%
  \hfill
  \begin{minipage}{0.48\columnwidth}
    \centering
    \includegraphics[width=\linewidth]{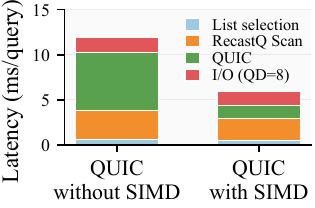}
    \vspace{-0.3in}
    \caption{QUIC with and without SIMD.}
    \label{fig:quic-simd-ablation}
  \end{minipage}
  \vspace{-0.1in}
\end{figure}

% Archived endpoint-by-endpoint update-scalability narration; superseded by
% the aggregate result below.
\begin{comment}
Figure~\ref{fig:update-scalability} starts every point from the same 100M-vector
checkpoint. Each \system point aggregates three successive 1M-vector
replacement rounds. \system scales from 2,800 replacement pairs/s on one core
to 74,434 on 72 cores (26.6$\times$), reducing aggregate 3M-replacement time
from 1,071.4 s to 40.3 s. SPFresh scales from 745 to 24,003 pairs/s
(32.2$\times$), while \system remains 3.76$\times$ and 3.10$\times$ faster at
the one- and 72-core endpoints, respectively. OdinANN peaks at 2,078 pairs/s
with 32 update cores and reaches 1,997 pairs/s at 72 cores, consistent with
its single background writer limiting parallel update throughput. Its one-
and four-core points use 125K and 500K replacements, respectively; the other
OdinANN points use 1M replacements and all rates include background-write
drain.
\end{comment}

Figure~\ref{fig:update-scalability-panel} reports the update throughput
by varying the number of threads.  We start from the same 100M-vector
index, then perform the same round of 1M deletes and 1M new inserts
for every worker count.  We see that \system scales the update
throughput by 26.6$\times$ from 1 to 72 workers and reaches 74,434
delete+insert pairs/s.  At 72 workers, it is 3.1--37.3$\times$ faster
than SPFresh and OdinANN. Note that all results include the work
required before the updated index can serve queries.

\section{Related Work}
\label{sec:related}

We have studied previous storage-optimized ANNS systems in
Section~\ref{sec:background}.  Since \system builds an in-memory index
that stores \recastq encodings, we discuss the related work on
in-memory ANNS indexes and quantization encodings in this section.

\Paragraph{In-memory ANNS indexes}
In-memory ANN spans tree and hashing methods such as KD-tree, ball tree,
VP-tree, FLANN, and LSH variants~\cite{KDTree,BallTree,VPTree,FLANN,LSH,
Multi_probe_LSH,LSH_forest}; proximity graphs such as HNSW and
NSG~\cite{RNG,FANNG,HNSW,tau_MNG,NSG}; and cluster-based indexes such as IVF,
IMI, ScaNN, and BLISS~\cite{IVF,IMI,ScaNN,Bliss}. Clustered layouts support
contiguous scans that combine naturally with quantization and SIMD execution.
Recast uses an IVF-compatible in-memory organization rather than introducing a
new coarse-search index; its focus is deciding from compact scan results which
exact vectors must subsequently be accessed from storage.

\Paragraph{Quantization encoding and reranking}
Compact vector representations include dimension-wise schemes such as
VA-File~\cite{VA_File}, ITQ~\cite{ITQ}, and LVQ~\cite{LVQ}; additive-codebook
methods such as AQ~\cite{AQ} and LSQ~\cite{LSQ}; and product methods such as
product quantization (PQ)~\cite{PQ} and optimized product quantization
(OPQ)~\cite{OPQ}. PQ and OPQ encode vectors into short Cartesian-product codes
and support asymmetric distance computation. PQ Fast Scan~\cite{PQfs},
Quicker ADC~\cite{QuickerADC}, and related kernels reorganize codes and lookup
tables for SIMD execution. These methods primarily optimize approximate
ranking, reconstruction, or scan throughput. RecastQ reuses a PQ4 execution
substrate but stores the magnitude and directional-error information needed for
refinement candidate selection.

RaBitQ~\cite{RabitQ} and Extended RaBitQ~\cite{ExtendedRabitQ} combine
randomized quantization with bounded distance estimators and lower-bound-guided
reranking. They are the closest encoding-level baselines to RecastQ. We use the
reference library and native reranking rule rather than replacing it with
Recast's selector. Recast differs in co-designing the uncertainty information,
the global Top-$K$ selector, page planning, and mutable external-store
architecture as one system.

% \Paragraph{Storage ANN architectures}

% Section~\ref{sec:background} describes the disk-graph and posting-list systems most closely related to Recast. Recast does not argue that graph traversal or posting retrieval is intrinsically slow. It instead studies a different external-storage role in which compact in-memory state completes refinement candidate selection before the storage tier supplies selected exact vectors.  This organization creates different construction, footprint, and maintenance tradeoffs from both disk graphs and storage-resident postings.

% \Paragraph{Dynamic vector indexes}

% Dynamic ANN work maintains search quality as vectors arrive and depart. The representative dynamic graph and posting mechanisms are summarized in Section~\ref{sec:background}. Recast differs in deriving batched local maintenance from packed in-memory lists and list-local exact-vector records, without maintaining graph connectivity or storage-resident posting replicas.

\section{Conclusion}
\label{sec:conclusion}

In conclusion, we have introduced a refinement-centered paradigm for
storage-optimized ANNS systems.  We have proposed and studied \system
as a concrete design of this paradigm.   It enjoys a simple storage
layout with low footprint and easy update support, and exploits a
compact in-memory IVF index with the \recastq encodings and the \quic
policy for efficient query processing.  We believe that our proposed
solution provides a promising answer to the central design question,
achieving high query throughput without sacrificing the recall and
index maintenance performance.

% By addressing query-dependent refinement demand, uncertainty quality, and in-memory processing cost, Recast realizes a refinement-centered storage ANN design in which compact in-memory information completes refinement candidate selection before exact-vector access. Its uncertainty-guided selector allocates access work at the Top-$K$ boundary, while RecastQ supplies compact magnitude and directional error information for that decision. The resulting system reaches graph-level high-recall throughput on SIFT100M, uses substantially less persistent storage, and preserves its query frontier under large stationary replacement churn.  These results suggest that compact encodings for storage ANN should be optimized not only for approximate ranking, but also for the exact-vector access decisions and operational structure they enable.

\bibliographystyle{ACM-Reference-Format}
\bibliography{references}
\endgroup

\clearpage
\appendix
\section{Proofs for Uncertainty-Guided Candidate Selection}
\label{app:proofs}

Fix a query and its scanned set $S=\{x_1,\ldots,x_n\}$, with
$n\geq K$, true distances $d_1\leq\cdots\leq d_n$, and fixed
tie-breaking. Define $R_{\mathrm{ref}}@K=K^{-1}\sum_{j=1}^K
\mathbf{1}\{L_j\leq\theta_K\}$. Expectations are conditional on
the true distances and scales. The results concern the uncapped
selector in Eq.~\ref{eq:selector}, not coarse-list misses or additional
candidate-cap truncation. We distinguish the true-distance spacing
needed for the recall bound from the endpoint spacing used for counts.

\subsection{Retention and a Gaussian Product Bound}
Let $N_j=\sum_{i\ne j}\mathbf{1}\{U_i<L_j\}$. Since $U_j\geq L_j$,
candidate $j$ is retained exactly when $N_j\leq K-1$. If all intervals
contain the true distances, an excluded candidate has
$d_j\geq L_j>\theta_K$, whereas at least $K$ candidates have
$d_i\leq U_i\leq\theta_K$. This proves
Theorem~\ref{thm:conditional-topk}, including ties.

For the probabilistic result, assume $\tau_i>0$ and
\begin{equation}
 \hat d_i=d_i+\tau_iZ_i,\qquad
 Z_i\stackrel{\mathrm{iid}}{\sim}\mathcal N(0,1).
 \label{eq:app-gaussian}
\end{equation}
Independence across candidates is an additional modeling assumption;
it does not follow from Gaussian marginal errors.

\begin{proposition}[Candidate-dependent recall bound]
\label{prop:app-product}
Under Eq.~\ref{eq:app-gaussian},
\begin{equation}
 \mathbb E[R_{\mathrm{ref}}@K]\geq\frac1K\sum_{j=1}^K
 \prod_{l=K+1}^n\Phi\!\left(
 \frac{d_l-d_j+z(\tau_l+\tau_j)}{\sqrt{\tau_l^2+\tau_j^2}}\right).
 \label{eq:app-product}
\end{equation}
\end{proposition}
\begin{proof}
For $j\leq K$, the event $A_j=\bigcap_{l>K}\{U_l\geq L_j\}$
implies retention: only the other $K-1$ true neighbors can contribute
to $N_j$. Conditional independence given $Z_j=t$ gives
\begin{align}
 g_{jl}(t)&=\Pr(U_l\geq L_j\mid Z_j=t)\notag\\
 &=\Phi\!\left(\frac{d_l-d_j+z(\tau_l+\tau_j)-\tau_jt}{\tau_l}\right),\\
 \Pr(A_j)&=\mathbb E\left[\prod_{l>K}g_{jl}(Z_j)\right].
\end{align}
These factors are nonnegative and nonincreasing. For any such $f,g$
and independent copies $T,T'$,
\begin{align}
 2\operatorname{Cov}(f(T),g(T))
 &=\mathbb E[(f(T)-f(T'))(g(T)-g(T'))]\notag\\
 &\geq0.
\end{align}
Products preserve this monotonicity, so induction yields
$\Pr(A_j)\geq\prod_{l>K}\Pr(U_l\geq L_j)$.
Since $\tau_lZ_l-\tau_jZ_j$ has Gaussian variance
$\tau_l^2+\tau_j^2$, each marginal equals the corresponding
$\Phi(\cdot)$ factor in Eq.~\ref{eq:app-product}. Averaging over
$j\leq K$ proves the result. This association argument is necessary:
crossings share $Z_j$ and are not unconditionally independent.
\end{proof}

\subsection{Integral Bound and Spacing Conditions}
Set $\tau_i=\tau>0$. Since $d_j\leq d_K$ for $j\leq K$,
Proposition~\ref{prop:app-product} gives
\begin{equation}
 \mathbb E[R_{\mathrm{ref}}@K]\geq\prod_{h=1}^{n-K}
 \Phi\!\left(\frac{d_{K+h}-d_K}{\sqrt2\tau}+\sqrt2z\right).
\end{equation}
Assume the \emph{true-distance} gap condition
$d_{K+h}-d_K\geq h\Delta_d$, where $\Delta_d>0$.
Let $a=\sqrt2z$, $b=\Delta_d/(\sqrt2\tau)$, and
$f(t)=\ln\Phi(a+bt)$. Because $f$ is negative and increasing,
right-endpoint Riemann sums imply
\begin{align}
 \ln\mathbb E[R_{\mathrm{ref}}@K]
 &\geq\sum_{h=1}^{n-K}f(h)
 \geq\int_0^{n-K}f(t)\,dt\notag\\
 &\geq\int_0^\infty f(t)\,dt
 =\frac{\sqrt2\tau}{\Delta_d}I(\sqrt2z).
\end{align}
The integral converges because $-\ln\Phi(x)\leq2(1-\Phi(x))$
for $x\geq0$ and the Gaussian tail is integrable. Exponentiation
establishes the expression in Theorem~\ref{thm:integral-recall-bound}
with $\Delta=\Delta_d$. For $n=K$, the empty product is one and
every scanned vector is retained.

\Paragraph{Distinguishing the two spacing models}
The preceding proof requires true-distance spacing $\Delta_d$, not
random upper-endpoint spacing $\Delta_U$. Independent nondegenerate
Gaussian errors do not preserve uniform spacing and cannot
simultaneously produce deterministic uniformly spaced endpoints.
Thus the recall expression is not established by endpoint spacing
alone; the main text's recall and count models must be read separately.

For the count model of Theorem~\ref{thm:numrefine}, suppose the sorted
upper endpoints have spacing $\Delta_U>0$ and a common scale $\tau$.
Retention is equivalent to $U_i\leq U_{(K)}+2z\tau$, so
\begin{equation}
 |C(S)|=\min\{n,K+\lfloor2z\tau/\Delta_U\rfloor\}.
\end{equation}
The main text's count applies before this expression reaches $n$.
Likewise, substituting empirical mean scales and gaps is illustrative,
not a guarantee: an observed endpoint gap is not a verified lower
bound on true-distance gaps. The example $\tau/\Delta=20.316$,
$z=2$ yields approximately $0.9803$ within the model, without
establishing its assumptions for every query.

\end{document}